\documentclass{aa}
\usepackage{natbib}
\usepackage{graphics}
\usepackage{amssymb}

\newcommand{\tpt}[1]{$\times 10^{#1}$}
\newcommand{\jypb}{Jy~beam$^{-1}$}
\newcommand{\kms}{km s$^{-1}$}
\newcommand{\htcop}{H$^{13}$CO$^+$}

\newcommand{\chtoh}{CH$_{3}$OH}
\newcommand{\cso}{C$^{17}$O}
\newcommand{\cao}{C$^{18}$O}

\newcommand{\cfs}{C$^{34}$S}
\newcommand{\hto}{H$_{2}$}
\newcommand{\nthp}{N$_2$H$^+$}
\newcommand{\hcop}{HCO$^+$}

\titlerunning{The NGC~1333-IRAS2 protostellar system on 500~AU scales}

\begin{document}

\title{The structure of the NGC~1333-IRAS2 protostellar system on
500~AU scales} 

\subtitle{An infalling envelope, a circumstellar disk, multiple
outflows, and chemistry}

\author{J.K. J{\o}rgensen\inst{1} \and M.R. Hogerheijde\inst{2} \and
E.F. van Dishoeck\inst{1} \and G.A. Blake\inst{3} \and
F.L. Sch\"{o}ier\inst{1,4}}

\authorrunning{J.K. J{\o}rgensen et al. }
\titlerunning{The structure of the NGC~1333-IRAS2 protostar}

\institute{Leiden Observatory, P.O. Box 9513, NL-2300 RA Leiden, The
Netherlands \and Steward Observatory, The University of Arizona, 933
N. Cherry Avenue, Tucson, AZ 85721-0065, USA \and Division of
Geological and Planetary Sciences, California Institute of Technology,
MS 150-21, Pasadena, CA 91125, USA \and Stockholm Observatory,
AlbaNova, SE-106 91 Stockholm, Sweden}

\offprints{Jes K.\, J{\o}rgensen} \mail{joergensen@strw.leidenuniv.nl}
\date{16 June 2003 / 3 October 2003}

\abstract{This paper investigates small-scale (500~AU) structures of
dense gas and dust around the low-mass protostellar binary
NGC~1333-IRAS2 using millimeter-wavelength aperture-synthesis
observations from the Owens Valley and
Berkeley-Illinois-Maryland-Association interferometers. The detected
$\lambda=3$~mm continuum emission from cold dust is consistent with
models of the envelope around \object{IRAS2A}, based on previously
reported submillimeter-continuum images, down to the 3\arcsec, or
500~AU, resolution of the interferometer data. Our data constrain the
contribution of an unresolved point source to 22~mJy. The importance
of different parameters, such as the size of an inner cavity and
impact of the interstellar radiation field, is tested. Within the
accuracy of the parameters describing the envelope model, the point
source flux has an uncertainty by up to 25\%. We interpret this point
source as a cold disk of mass $\gtrsim 0.3$~$M_\odot$. The same
envelope model also reproduces aperture-synthesis line observations of
the optically thin isotopic species C$^{34}$S and H$^{13}$CO$^+$. The
more optically thick main isotope lines show a variety of components
in the protostellar environment: N$_2$H$^+$ is closely correlated with
dust concentrations as seen at submillimeter wavelengths and is
particularly strong toward the starless core \object{IRAS2C}. We
hypothesize that N$_2$H$^+$ is destroyed through reactions with CO
that is released from icy grains near the protostellar sources IRAS2A
and B. CS, HCO$^+$, and HCN have complex line shapes apparently
affected by both outflow and infall. In addition to the east-west jet
seen in SiO and CO originating from IRAS2A, a north-south velocity
gradient near this source indicates a second, perpendicular
outflow. This suggests the presence of a binary companion within
$0{\farcs}3$ (65~AU) from IRAS2A as driving source of this
outflow. Alternative explanations of the velocity gradient, such as
rotation in a circumstellar envelope or a single, wide-angle
($90^\circ$) outflow are less likely.

} 

\maketitle

\section{Introduction\label{s:intro}}
Our understanding of the cloud cores that form stars has benefited
significantly from the advent over the last years of
(sub)millimeter-continuum bolometer cameras. Sensitive, spatially
resolved measurements have allowed quantitative testing of models of
starless/pre-stellar cores and envelopes around young stars
\citep[e.g.][]{shirley00, shirley02, hogerheijde00sandell, motte01,
jorgensen02, schoeier02, belloche02}. Not only do these models sketch
the evolution of the matter distribution during star formation, they
also can serve as `baselines' for interpreting higher resolution
observations obtained with millimeter interferometry. Such data
address the presence and properties of circumstellar disks during the
early, embedded phase \citep[e.g.][]{hogerheijde98, hogerheijde99,
looney00}. This Paper presents millimeter aperture-synthesis
observations of continuum and line emission of the young protobinary
system \object{NGC~1333-IRAS2}, and uses modeling results based on
single-dish submillimeter continuum imaging from
\citeauthor{jorgensen02}~(2002;~Paper~I
hereafter)\defcitealias{jorgensen02}{Paper~I} to interpret the data in
terms of a collapsing envelope, a disk, and (multiple) outflows on
500~AU scales.

\begin{figure*}
\resizebox{\hsize}{!}{\includegraphics{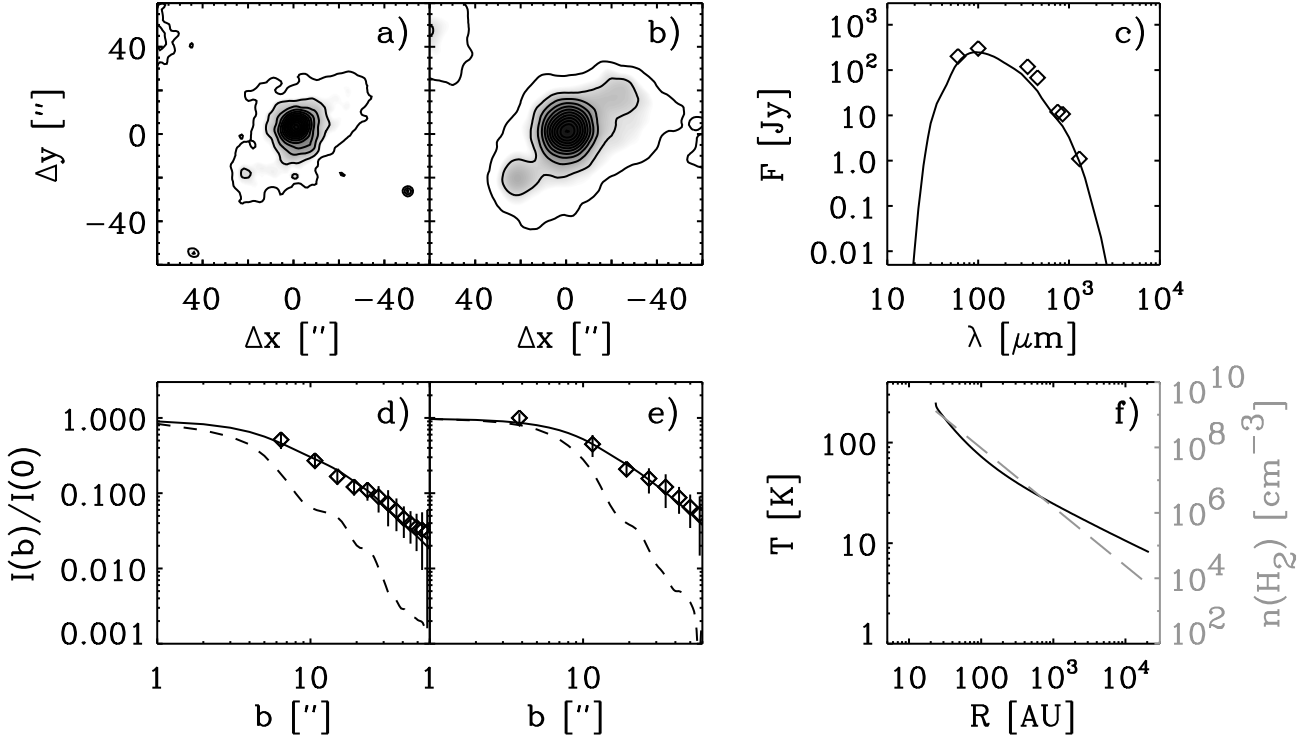}}
\caption{a) and b) SCUBA maps of \object{IRAS2} at 450 and 850~$\mu$m,
respectively, centered on \object{IRAS2A}. c) Observed SED (symbols) and model
fit (solid curve). d) and e) Brightness profiles at 450 and 850~$\mu$m
(symbols) and the model fit (solid curve). The dashed curves show the
beam profiles. f) Density (dashed line) and temperature (solid line)
distributions of best-fit model. See Paper~I for
details.}\label{n1333i2paper1}
\end{figure*}

The deeply embedded \citep[`class 0';][]{lada87,andre93} young stellar
system \object{NGC~1333-IRAS2} (\object{IRAS 03258+3104}; hereafter
\object{IRAS2}) has been the subject of several detailed studies. It
is located in the \object{NGC~1333} molecular cloud, well known for
harboring several class~0 and I objects, and was first identified from
IRAS data by \cite{jennings87}. Quoted distances to \object{NGC~1333}
range from 220~pc \citep{cernis90} to 350~pc \citep{herbig83}; here we
adopt 220~pc in accordance with \citetalias{jorgensen02}. At this
distance the bolometric luminosity of \object{IRAS2} is
$16~L_\odot$. Submillimeter-continuum imaging \citep[][and
Fig.~\ref{n1333i2paper1} below]{sandell01} and high-resolution
millimeter interferometry \citep{blake96,looney00} have shown that
\object{IRAS2} consists of at least three components: two young
stellar sources 2A and, 30\arcsec\ to the south-east, 2B; and one
starless condensation 2C, 30\arcsec\ north-west of 2A. The sources 2A
and 2B are also detected at cm wavelengths
\citep{rodriguez99,reipurth02}.

Maps of CO emission of the \object{IRAS2} region show two outflows,
directed north-south and east-west \citep{liseau88, sandell94knee,
knee00, engargiola99}. Both flows appear to originate to within a few
arcseconds from 2A \citep{engargiola99}, indicating this source is a
binary itself although it has not been resolved so-far. The different
dynamical time scales of both flows suggests different evolutionary
stages for the binary members, which lead \cite{knee00} to instead
propose 2C (30\arcsec\ from 2A) as driving source of the north-south
flow. It is unclear how well dynamic time scales can be estimated for
outflows that propagate through dense and inhomogeneous clouds such as
\object{NGC~1333}. Single-dish CS and HCO$^+$ maps also show
contributions by the outflow, especially for CS
\citep{wardthompson01}. The north-south outflow may connect to an
observed gradient in centroid velocities near 2A, but the authors
cannot rule out rotation in an envelope perpendicular to the east-west
flow.

\citetalias{jorgensen02} determined the physical properties of the
\object{IRAS2} envelope using one-dimensional radiative transfer
modeling of {\it Submillimeter Common User Bolometer Array\/} (SCUBA)
maps and the long-wavelength spectral energy distribution
(SED). Assuming a single radial power-law density distribution,
$\rho\propto r^{-p}$, an index $p=1.8$ and a mass of 1.7~$M_\odot$
within 12,000~AU was found (see Table \ref{paper1params} and
Fig.~\ref{n1333i2paper1}). Monte-Carlo modeling of the molecular
excitation and line formation of \cao\ and \cso\ observations yield a
CO abundance of $2.6\times 10^{-5}$ with respect to H$_2$, a factor
4-10 lower than what is found in local dark clouds
\citep[e.g.][]{frerking82,lacy94}.

\begin{table}
\caption{The parameters for \object{IRAS2} from
\citetalias{jorgensen02}.}\label{paper1params}
\begin{tabular}{ll}\hline\hline
Distance, $d$                        & 220~pc \\
$L_{\rm bol}$                        & 16~$L_\odot$ \\
$T_{\rm bol}$                        & 50~K \\ \hline
\multicolumn{2}{l}{\sl Envelope parameters:} \\ \hline
Inner radius ($T=250$~K), $R_i$      & 23.4 AU       \\
Outer radius, $R_{\rm 10 K}$$^{a}$ & 1.2\tpt{4} AU \\
Density at 1000~AU, $n$(\hto)        & 1.5\tpt{6} cm$^{-3}$ \\
Slope of density distribution, $p$   & 1.8           \\
Mass, $M_{\rm 10 K}$$^{a}$         & 1.7 $M_\odot$ \\
CO abundance, [CO/\hto]              & 2.6\tpt{-5}   \\\hline
\end{tabular}

Notes: $^{a}$The outer boundary is not well constrained, but taken
to be the point where the temperature in the envelope has dropped to
10~K. The mass refers to the envelope mass within this radius.
\end{table}

This Paper presents $\lambda=3$~mm interferometric observations of
\object{IRAS2} in a range of molecular emission lines probing dense
gas and continuum emission tracing cold dust. It builds on the
modeling of Paper~I by using it as a framework to interpret the
small-scale structure revealed by the aperture-synthesis data. Section
\ref{observations} describes the observations and reduction
methods. Section \ref{continuum} analyzes the continuum emission, and
compares it to the previously derived models. Section \ref{lines}
presents the molecular-line maps and discusses the physical and
chemical properties of the gas in the proximity of
\object{IRAS2}. Section \ref{velgradient} reports a pronounced
north-south velocity gradient around \object{IRAS2A} and explores
rotation or outflow as possible explanations. Section \ref{conclusion}
concludes the Paper by summarizing the main findings. A companion
paper \citep{i2art} presents a detailed study of the bow shock at the
tip of the east-west jet from \object{IRAS2} based on single-dish and
interferometric (sub)millimeter observations.

\section{Observations\label{observations}}
\subsection{Interferometer data\label{ss:interferometer}}
\object{IRAS2} ($\alpha(2000)=03^{\rm h}28^{\rm m}56\fs29$;
$\delta(2000)=31^\circ14\arcmin33\farcs93$) was observed with the
Millimeter Array of the Owens Valley Radio Observatory
(OVRO)\footnote{The Owens Valley Millimeter Array is operated by the
California Institute of Technology under funding from the US National
Science Foundation.} between October 5, 1994 and January 1, 1995 in
the six-antenna L- and H-configurations. Tracks were obtained in two
frequency settings at 86 and 97~GHz, and each track observed
alternatingly two fields: the source positions discussed in this Paper
and the bow shock at the end of the eastern jet \citep{i2art}. The
observed tracks cover projected baselines of 3.1-70~k$\lambda$ at
86~GHz. The observed lines are listed in Table~\ref{obs_overview}, and
were recorded in spectral bands with widths of 32~MHz ($\sim
100$~km~s$^{-1}$). \htcop\ $1-0$ and CS $2-1$ were observed in 128
spectral channels and the remaining lines in 64 spectral channels. The
complex gain variations were calibrated by observing the nearby
quasars 0234+285 and 3C84 approximately every 20 minutes. Fluxes were
calibrated by observations of Uranus and Neptune. Calibration and
flagging of visbilities with clearly deviating amplitudes and/or
phases was performed with the MMA reduction package
\citep{scoville93}.

The millimeter interferometer of the Berkeley-Illinois-Maryland
Association (BIMA)\footnote{The BIMA array is operated by the
Universities of California (Berkeley), Illinois, and Maryland, with
support from the National Science Foundation.} observed \object{IRAS2}
on November 4-5, 2000, and January 20-21, February 20, and June 5-6,
2001. The array B-, C-, and D-configurations provided projected
baselines of 1.7--68~k$\lambda$. The lines of HCO$^+$ $J$=1--0, HCN
1--0, N$_2$H$^+$ 1--0, and C$^{34}$S 2--1 were recorded in 256-channel
spectral bands with a total width of 6.25~MHz ($\sim
20$~km~s$^{-1}$). The complex gain of the interferometer was
calibrated by observing the bright quasars 3C84 (4.2~Jy) and 0237+288
(2.3~Jy) approximately every 20 minutes. The absolute flux scale was
bootstrapped from observations of Uranus. The rms noise levels are
0.14~Jy~beam$^{-1}$ in the 24~kHz channels, with a synthesized beam
size of $8.2''\times 7.5''$ FWHM. The data were calibrated with
routines from the MIRIAD software package \citep{sault95}.

In the reduction, data points with clearly deviating phases or
amplitudes were flagged. The maps were cleaned down to 3 times the rms
noise using the MIRIAD `clean' routine. The strong continuum of the
two central point sources allowed self-calibration, which was applied
and used to correct the spectral line data. The naturally weighted
continuum observations typically had rms noise better than 1\tpt{-3}
\jypb\ with half power beam widths (HPBW) of $\approx$ 3\arcsec\ for
the OVRO observations and $\approx$ 8\arcsec\ for the BIMA data (see
Table~\ref{cont_source_table}). Table~\ref{obs_overview} lists the
details of the line observations.
\begin{table}
\caption{Line data of \object{IRAS2} discussed in this Paper.}\label{obs_overview}
\begin{tabular}{llllll}\hline\hline
Molecule & Line & Rest freq.   & Observed with    \\ \hline
CH$_3$OH   & $2_1-1_1$         & \phantom{1}97.5828 & OVRO \\[0.5ex]
CS         & $2-1$             & \phantom{1}97.9810 & OSO, OVRO \\
           & $3-2$             & 146.9690           & IRAM 30m \\
           & $5-4$             & 244.9356           & IRAM 30m, JCMT$^{b}$ \\
           & $7-6$             & 342.8830           & JCMT$^{b}$ \\[0.5ex]
C$^{34}$S  & $2-1$             & \phantom{1}96.4129 & IRAM 30m, BIMA \\
           & $5-4$             & 241.0161           & JCMT \\[0.5ex]
HCN        & $1-0$$^{a}$       & \phantom{1}88.6318 & OSO, BIMA \\[0.5ex]
\htcop\    & $1-0$             & \phantom{1}86.7543 & OSO, OVRO \\[0.5ex]
HCO$^+$    & $1-0$             & \phantom{1}89.1885 & OSO, BIMA \\[0.5ex]
N$_2$H$^+$ & $1-0$$^{a}$       & \phantom{1}93.1737 & OSO, BIMA \\[0.5ex]
SiO        & $2-1$             & \phantom{1}86.8470 & OVRO \\[0.5ex]
SO         & $2_2-1_1$         & \phantom{1}86.0940 & OVRO \\[0.5ex]
SO$_2$     & $7_{3,5}-8_{2,6}$ & \phantom{1}97.7023 & OVRO \\\hline
\end{tabular}

Notes: $^{a}$Hyperfine splitting observed in one
setting. $^{b}$Archival data.
\end{table}

\subsection{Single-dish data\label{ss:singledish}}
Using the Onsala 20~m telescope (OSO)\footnote{The Onsala 20~m
telescope is operated by the Swedish National Facility for Radio
Astronomy, Onsala Space Observatory at Chalmers University of
Technology.} a number of molecules were observed towards
\object{IRAS2A} in March 2002. These are listed in
Table~\ref{obs_overview}. We also include line data from
\cite{jorgensen02,paperii} obtained with the IRAM~30m\footnote{The
IRAM 30~m telescope is operated by the Institut de Radio Astronomie
Millim\'etrique, which is supported by the Centre National de
Recherche Scientifique (France), the Max Planck Gesellschaft (Germany)
and the Instituto Geogr\'afico Nacional (Spain).}  and James Clerk
Maxwell (JCMT)\footnote{The JCMT is operated by the Joint Astronomy
Centre in Hilo, Hawaii on behalf of the parent organizations PPARC in
the United Kingdom, the National Research Council of Canada and The
Netherlands Organization for Scientific Research} telescopes, and
spectra taken from the JCMT archive\footnote{The JCMT archive at the
Canadian Astronomy Data Centre is operated by the Dominion
Astrophysical Observatory for the National Research Council of
Canada's Herzberg Institute of Astrophysics.}. The lines were
converted to the main beam antenna temperature scales using the
appropriate efficiencies, and low-order polynomial baselines were
fitted and subtracted.

\section{The continuum emission\label{continuum}}
\begin{figure}
\resizebox{\hsize}{!}{\includegraphics{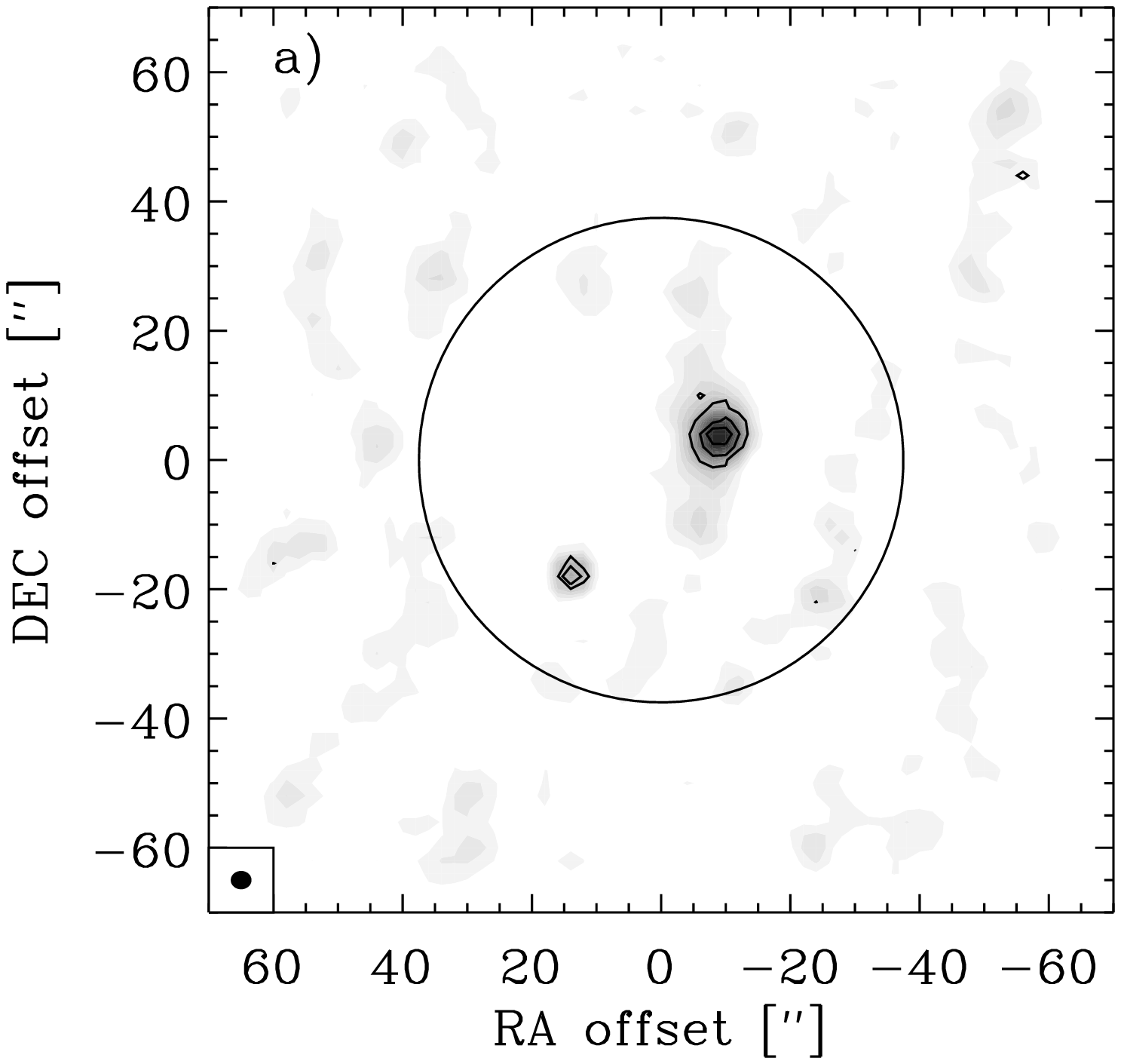}\includegraphics{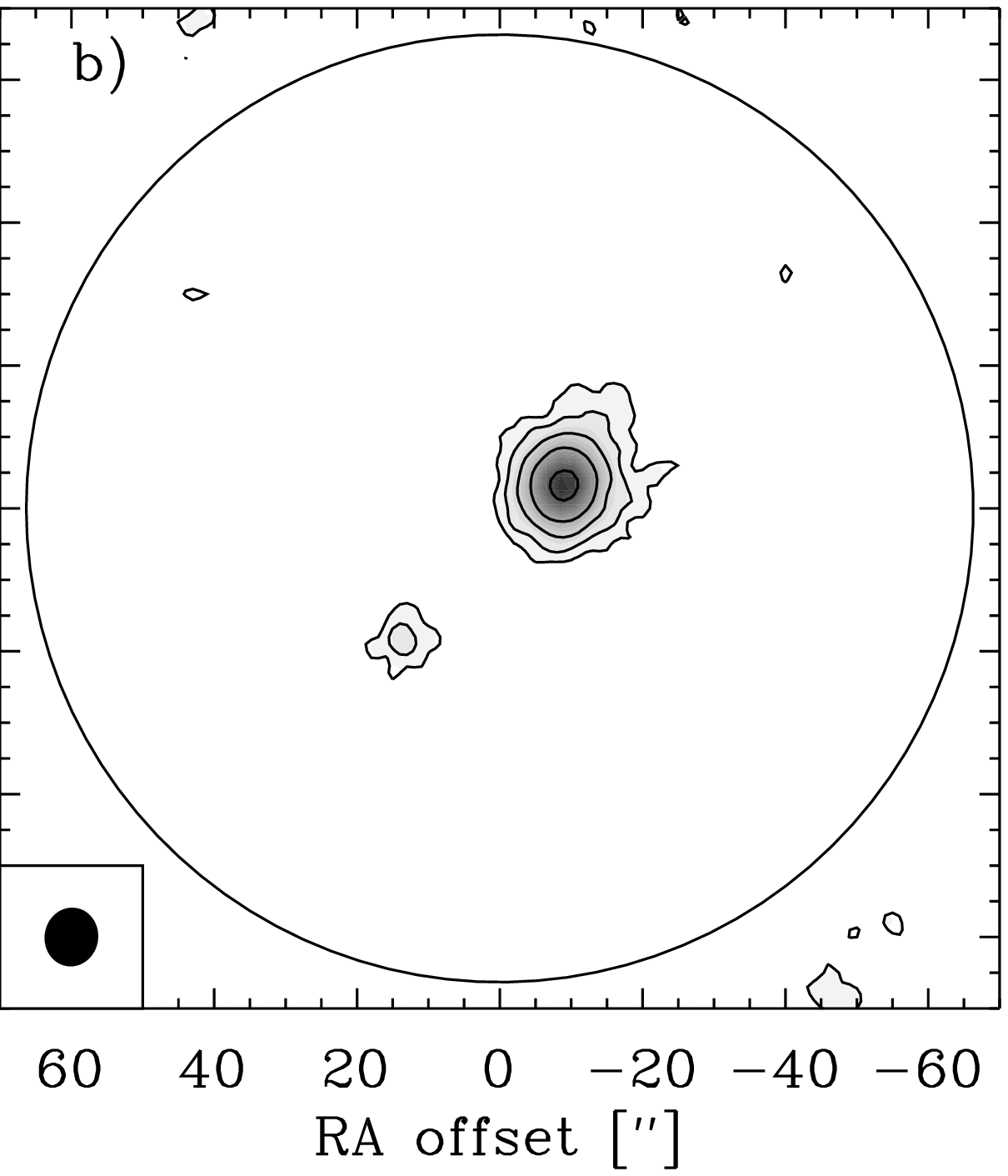}}
\caption{Maps of continuum emission at 86--89~GHz from OVRO (a) and
BIMA (b). Offsets are with respect to the pointing center of
$\alpha(2000)=03^{\rm h}28^{\rm m}56\fs29$ and
$\delta(2000)=31^\circ14\arcmin33\farcs93$. Contours are shown at
3$\sigma$, 6$\sigma$, 12$\sigma$, 24$\sigma$ and 48$\sigma$, where
$\sigma$ is the RMS noise level of Table~\ref{cont_source_table}. The
filled ellipses in the lower left corner of the panels indicate the
synthesized beam sizes; the large circles show the 50\% sensitivity
levels of the primary beams.}\label{cont_source}
\end{figure}

The 3~mm continuum images clearly show the two components 2A at
($-9''$, $+3''$) and 2B at ($+14''$, $-18''$)
(Fig.~\ref{cont_source}). Table \ref{cont_source_table} lists the
results of fits of two circular Gaussians to the visibility
data. Consistent with \cite{looney00}, 2A is the stronger of the
two. Differences in detected fluxes between the OVRO and BIMA data
sets indicate that the emission is extended, and varying amounts are
picked up by the respective $(u,v)$ coverages of the arrays. Emission
from the third source 2C, north-west of 2A, is not detected,
supporting the suggestion that it has not yet formed a star and lacks
a strong central concentration.

\begin{table}
\caption{Results of fits to the visibilities.
\label{cont_source_table}}
\begin{tabular}{lll}\hline\hline
                       & OVRO                 & BIMA \\ \hline
RMS (\jypb)            & 1.0\tpt{-3}          & 0.9\tpt{-3}  \\
Beam                   & 3.2\arcsec$\times$2.8\arcsec & 8.2\arcsec$\times$7.5\arcsec \\ \hline
                       & \multicolumn{2}{c}{\object{IRAS2A}}          \\ \hline
$F_{\rm tot}$ (Jy)     & $0.035$      & $0.040$  \\
X-offset (\arcsec)     & $-9.17$      & $-8.92$  \\
Y-offset (\arcsec)     & $3.67$       & $3.10$   \\ \hline
                       & \multicolumn{2}{c}{\object{IRAS2B}}          \\ \hline
$F_{\rm tot}$ (Jy)     & $0.012$      & $0.014$  \\
X-offset (\arcsec)     & $13.66$      & $13.78$  \\
Y-offset (\arcsec)     & $-17.51$     & $-18.37$ \\ \hline
\end{tabular}

Notes: \object{IRAS2A} is marginally resolved, whereas \object{IRAS2B} is unresolved.
\end{table}

\subsection{A model for the continuum emission\label{analysiscont}}
\begin{figure}
\resizebox{\hsize}{!}{\includegraphics{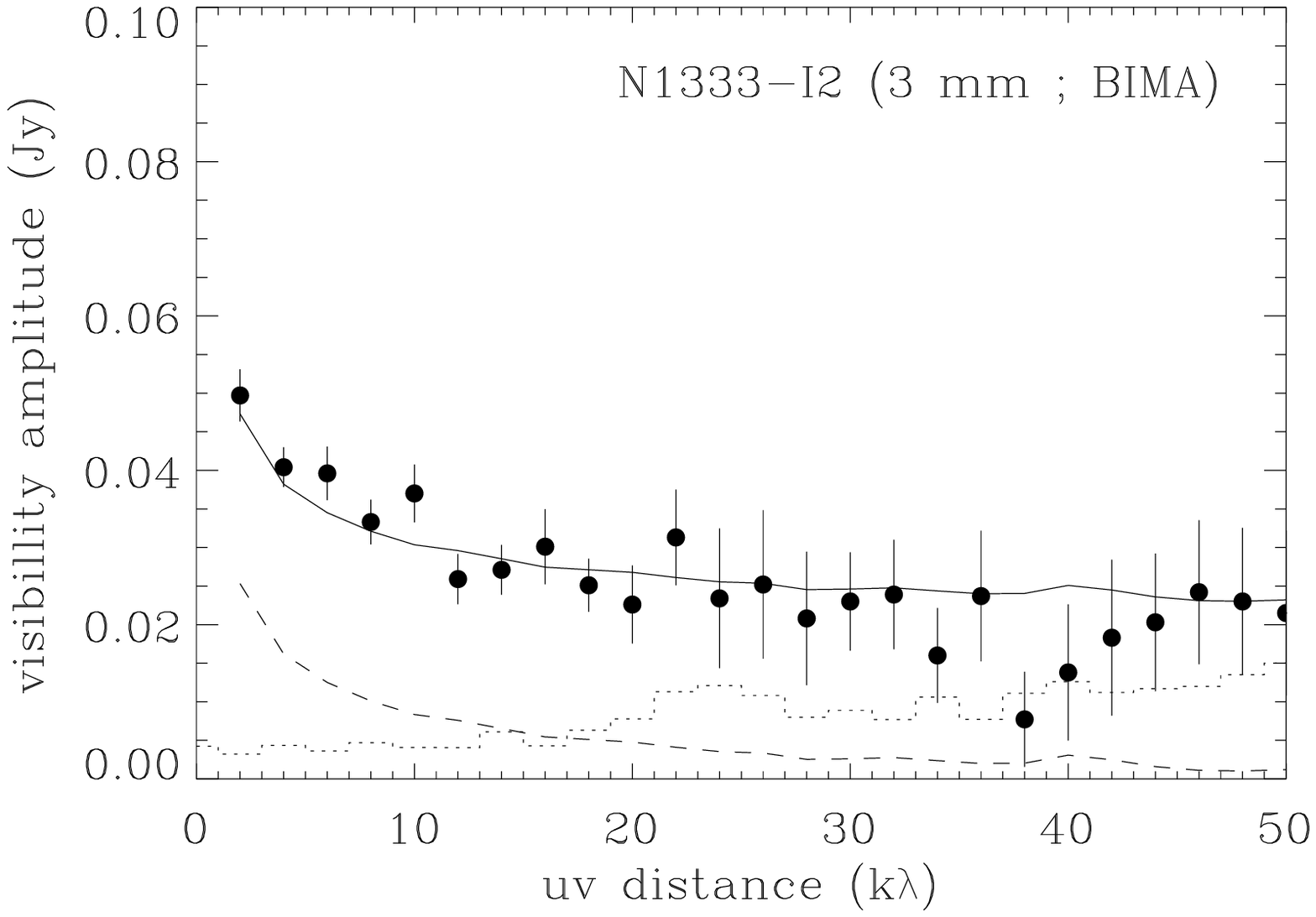}}
\resizebox{\hsize}{!}{\includegraphics{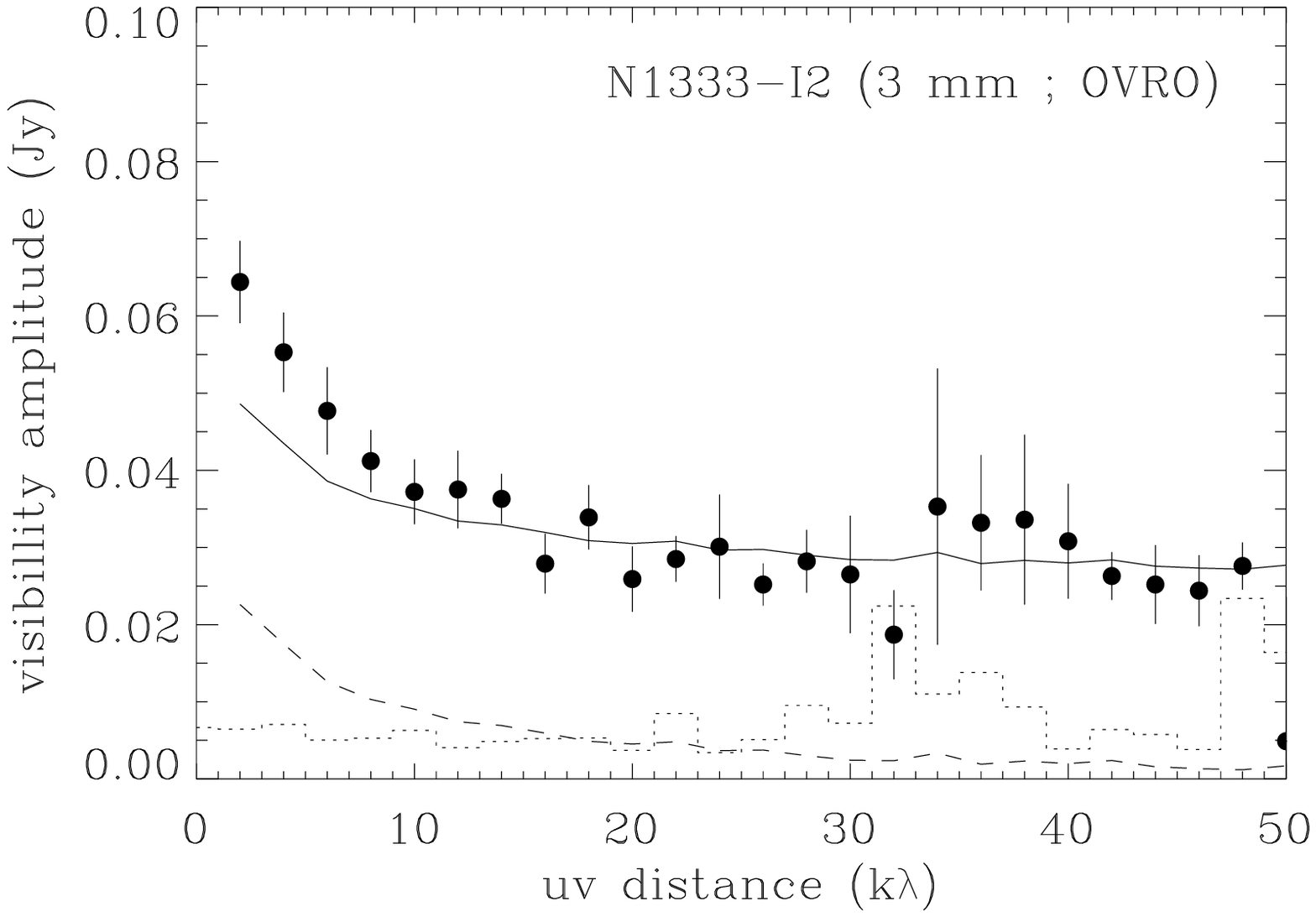}}
\caption{Visibility amplitudes of the observed continuum emission from
the BIMA (upper panel) and OVRO (lower panel) observations as a
function of projected baseline length in k$\lambda$ centered at the
position of \object{IRAS2A}. The data are plotted as filled symbols
with 1$\sigma$ error bars, the continuous lines indicate the
predictions from the continuum model of \citetalias{jorgensen02} with
the same $(u,v)$ sampling as the observations - respectively with
(solid) and without (dashed) an unresolved compact source of 22~mJy
added to the model. The dotted histogram indicates the
zero-expectation level: the expected amplitude signal due to noise
alone in the absence of source emission.}\label{uvsamcont}
\end{figure}

The different detected fluxes from OVRO and BIMA in
Table~\ref{cont_source_table} and comparison of the interferometer
images of Fig.~\ref{cont_source} and the SCUBA images of
Fig.~\ref{n1333i2paper1} clearly show that the arrays have resolved
out significant amounts of extended emission because of their limited
$(u,v)$ coverage. The envelope model (density, temperature, dust
emissivity as function of wavelength) by \citetalias{jorgensen02}
predicts sky-brightness distributions at 3~mm, and fluxes in the
interferometer beams after sampling at the actual $(u,v)$ positions
and subtracting the contribution from 2B. This latter subtraction of
the Gaussian fit to 2B only affects the results minimally, indicating
that 2B is well separated from and much weaker than
2A. Fig.~\ref{uvsamcont} compares the predicted flux as function of
projected baseline length with the data. In addition to the model
envelopes, we have included as free parameter the flux of an
unresolved point source ($<3\arcsec$). Because a point source
contributes equally on all baselines, this addition corresponds to a
vertical offset of the model curve in the plots. Such offsets are
apparent in both OVRO and BIMA data, and a point source flux of 22~mJy
at 3~mm provides an adequate fit to the data when added to the model
envelopes.

The inferred point source flux is consistent with \cite{looney03}, who
find a 20~mJy source at 2.7~mm associated with \object{IRAS2A}. For reasonable
assumptions about the spectral index of the point source
($\alpha$=2--4 if thermal) this component does not contribute
significantly to the flux in the much larger SCUBA beams, and
therefore does not invalidate the SCUBA-based envelope models. The
thermal nature of the point source is supported by detection of 2A at
a flux of 0.22~mJy at 3.6~cm and a resolution of 0.3\arcsec\ with the
VLA \citep{reipurth02}. This yields a spectral index of 1.9 between
3.6~cm and 3.3~mm, consistent with optically thick thermal emission. A
similar conclusion was reached by \cite{rodriguez99} based on the
spectral index from VLA observations of \object{IRAS2A} at 3.6 and 6~cm.

Assuming that the point source emission is optically thin and thermal,
the inferred flux of 22~mJy corresponds to a dust mass of $3.3\times
10^{-3}$~$M_\odot$ if we adopt an average temperature of 30~K and a
emissivity per unit (dust) mass at 3.5~mm of
$\kappa=0.24$~cm$^{2}$~g$^{-1}$ from extrapolation of the opacities by
\cite{ossenkopf94} for grains with thin ice mantles as was assumed in
the envelope models in \citetalias{jorgensen02}. With a standard
gas-to-dust ratio of 100, the total mass is 0.33~$M_\odot$. If the
emission is optically thick as the spectral index indicates, this is
in fact a lower limit to the mass. The favored explanation for this
compact mass distribution is a circumstellar disk.

\subsection{Parameter dependency of the continuum model}
\begin{figure*}
\resizebox{\hsize}{!}{\includegraphics{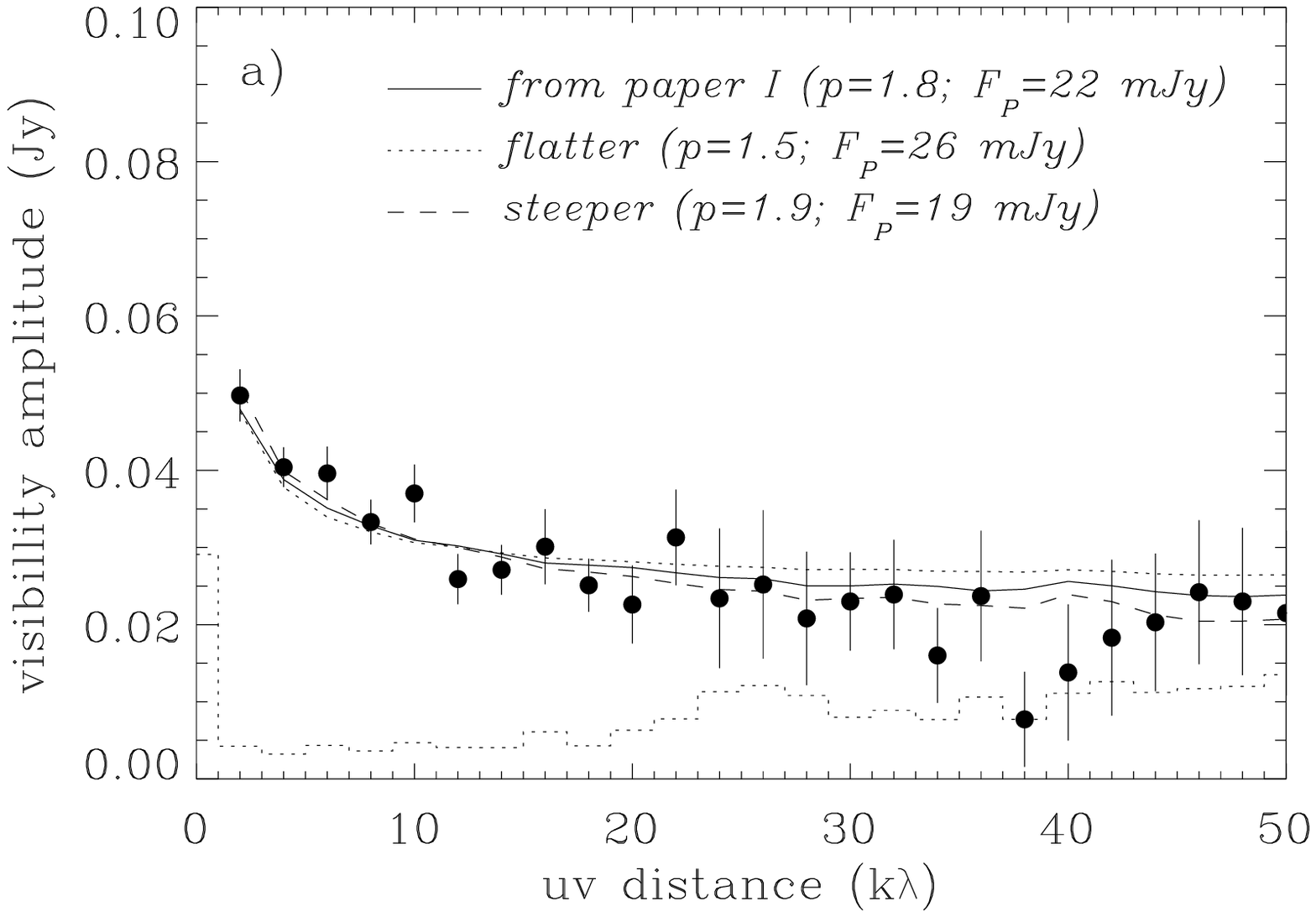}\includegraphics{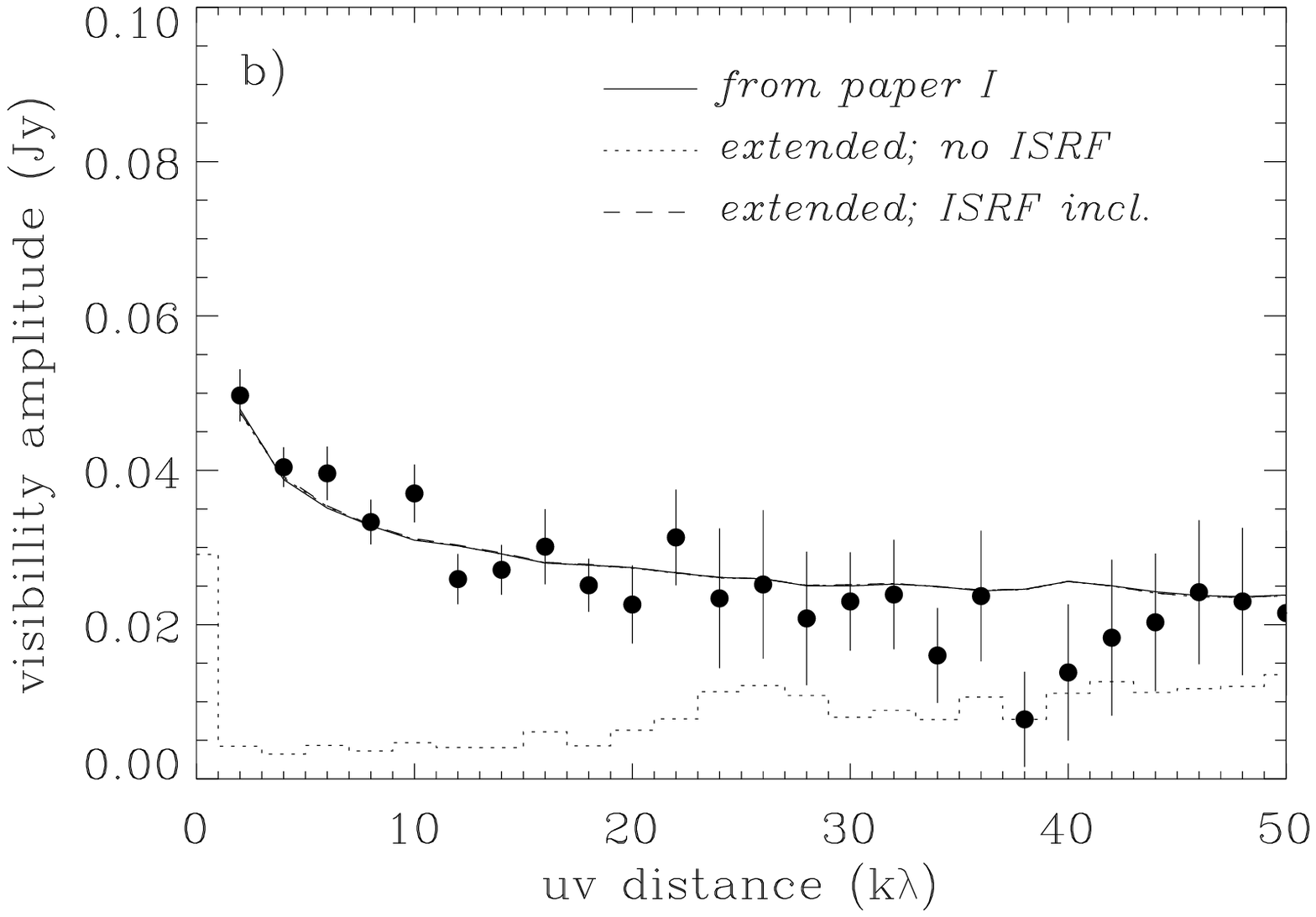}}
\resizebox{\hsize}{!}{\includegraphics{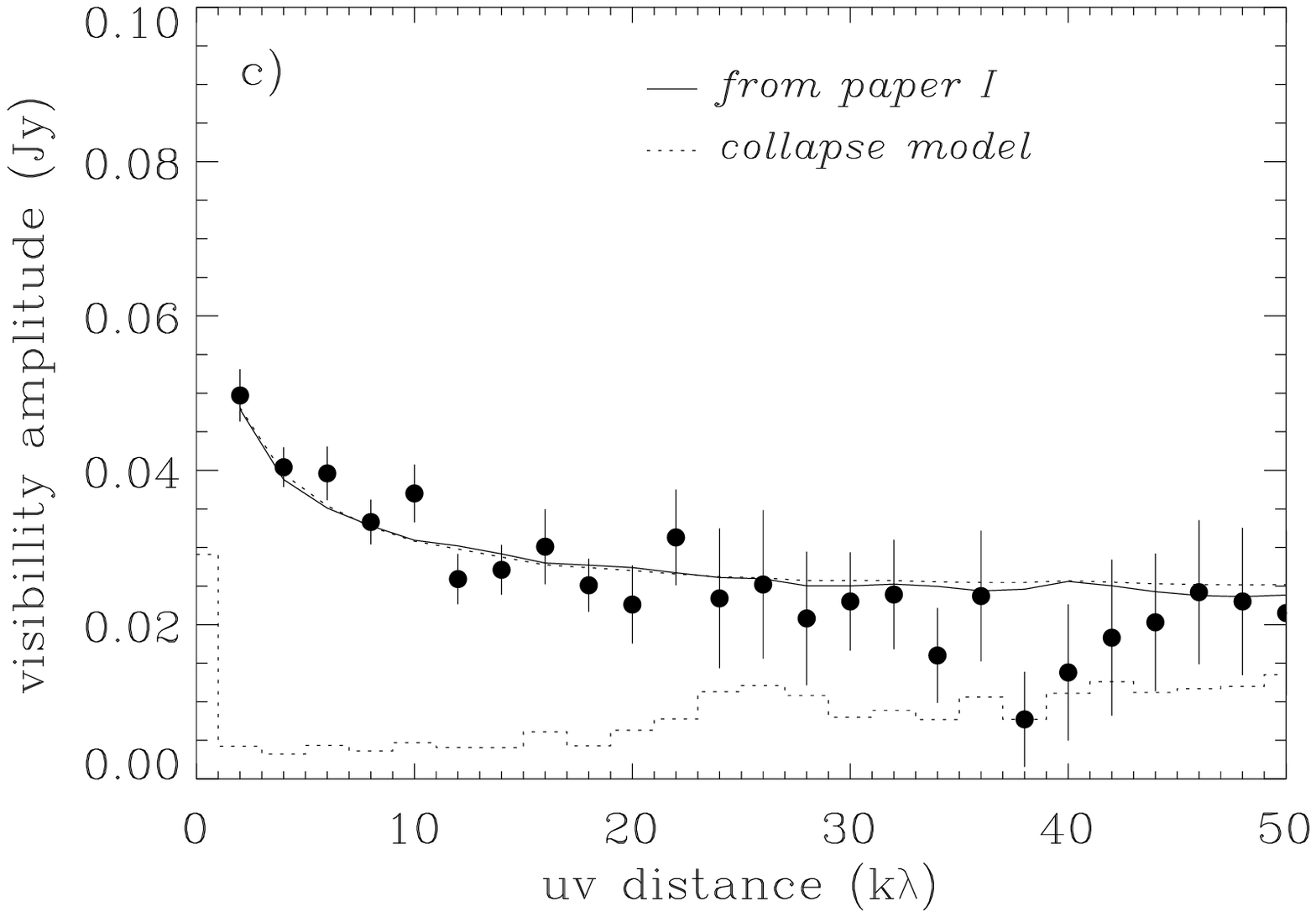}\includegraphics{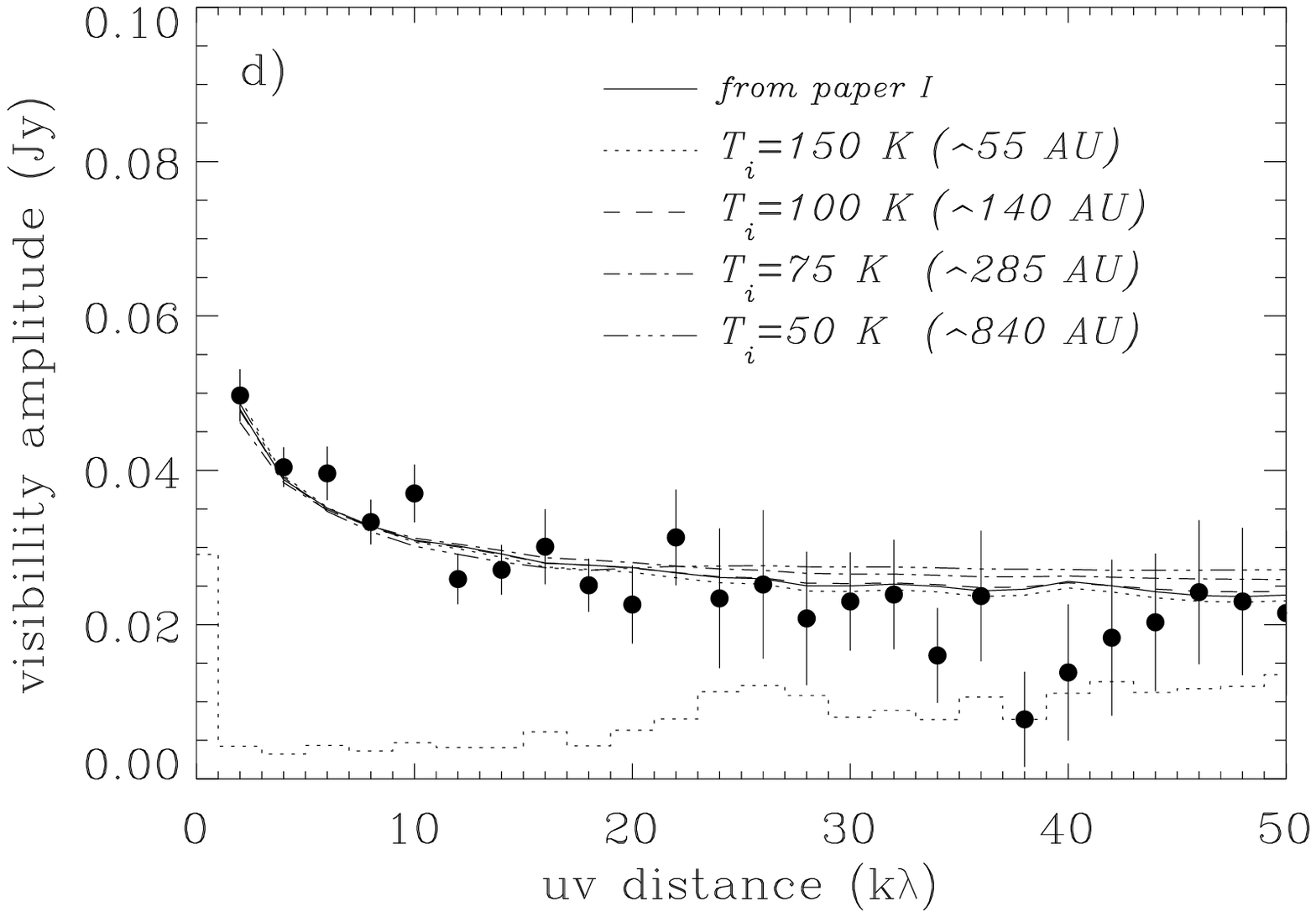}}
\caption{Visibility amplitudes of the observed BIMA continuum emission
as in Fig.~\ref{uvsamcont} compared to various input models centered
at the position of \object{IRAS2A}. \emph{Upper panels:} a) models
with changing steepness of the density profile. b) test of different
values of the outer radius and inclusion of the interstellar radiation
field. Models with an outer radius 3 times larger than the model from
paper I (i.e. 36000~AU) are shown. \emph{Lower panels:} c) fit to the
inside-out collapse model of \cite{shu77} with parameters constrained
independently by molecular line observations and SCUBA continuum
observations. d) models with changing size of the inner
radius.}\label{bima_uvmodels}
\end{figure*}

To test the validity of the envelope model a number of parameters were
varied within the constraints set by the modeling of the SCUBA
observations (Fig.~\ref{bima_uvmodels}). The uncertainty in the
power-law index from the SCUBA model of Paper~I is $\pm 0.2$. Over
this range of density slopes we find central point source fluxes of
$22\pm 4$~mJy, with a clear degeneracy between the slope of the
density profile and the flux of the central point source. This is
similar to what \cite{harvey03} find in a detailed analysis of
high-resolution millimeter continuum observations of the class~0
object \object{B335}. Both BIMA and OVRO data sets are fitted well
within the uncertainties using the density profile slope from the
SCUBA data, although the actual best fit model to the OVRO data has a
slightly steeper density slope ($p=1.9$) and a lower point source
flux. The interferometry data cannot constrain the slope of the
density profile further than its uncertainty from the SCUBA model. So
although the data can be fitted with a single power-law density
envelope from the scales probed by the SCUBA observations down to the
scales probed by the interferometry observations, a steepening or
flattening of the density profile at small scales cannot be ruled out.

\begin{figure}
\resizebox{\hsize}{!}{\includegraphics{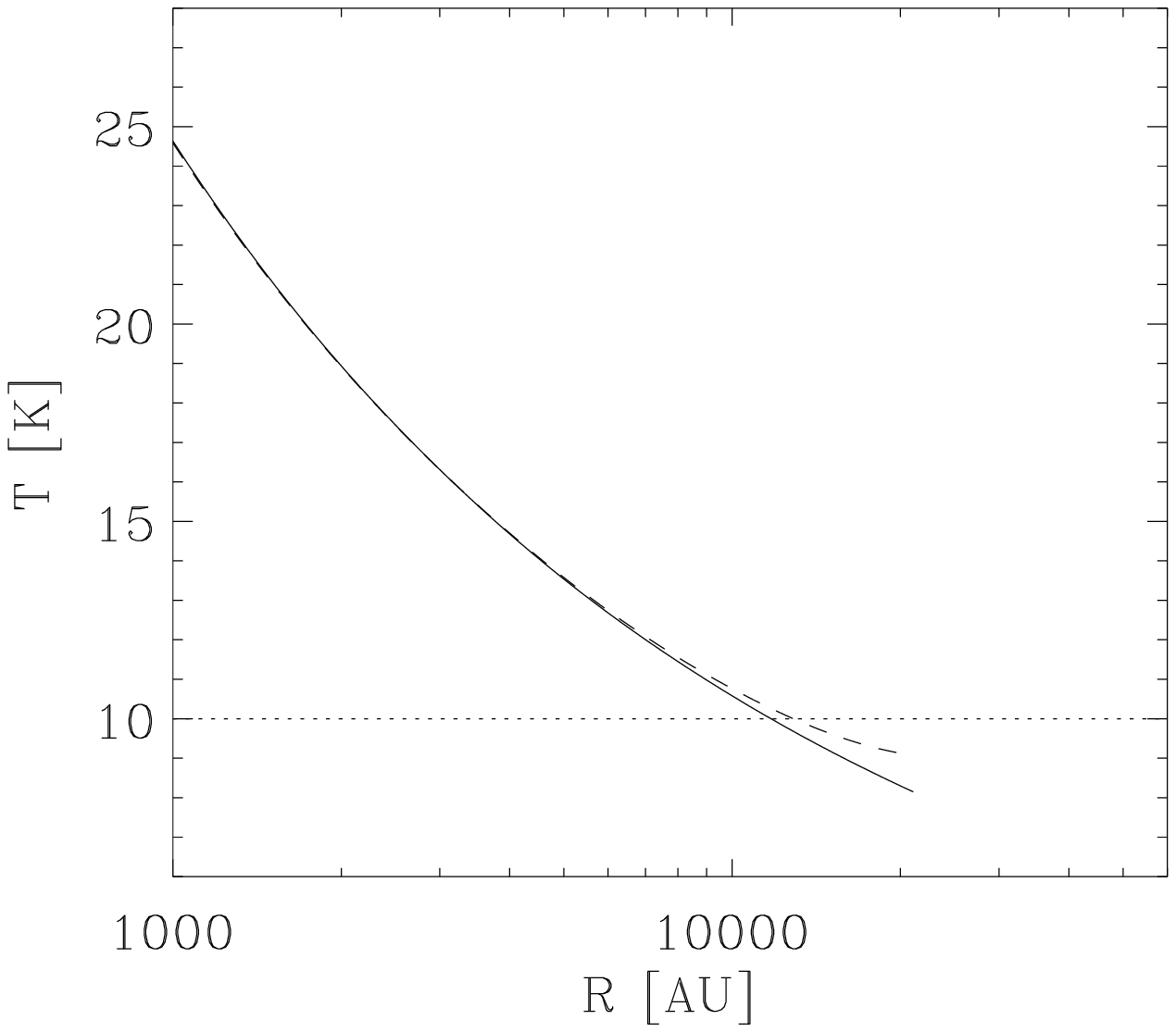}}
\caption{Temperature profile in the outermost region of the envelope
without (solid line) and with (dashed line) contributions from the
interstellar radiation field. The dotted line indicates the
temperature of 10~K corresponding to the envelope outer radius
\citepalias{jorgensen02}.}\label{extheating}
\end{figure}

\cite{harvey03} find that the uncertainty in the central point source
flux dominates over uncertainties in other model parameters such as
external heating by the interstellar radiation field (ISRF),
wavelength dependence of the dust emissivity, outer radius of the
envelope, and deviations from spherical symmetry (e.g., an evacuated
outflow cavity). Because our interferometer data only sample the inner
regions, they are not sensitive to variations in the outer radius or
inclusion of heating by the ISRF. The latter hardly affects the
temperature structure because the source is relatively luminous and
dominates the heating as is seen in Fig.~\ref{extheating}.

In similar studies to that presented in \citetalias{jorgensen02},
\cite{shirley02} and \cite{young03} modeled the SEDs and brightness
profiles from SCUBA observations of protostellar sources using 1D
radiative transfer, assuming power-law density profiles and solving
for the temperature structure. Two differences exist, however, in the
approaches taken in these two papers and our \citetalias{jorgensen02}:
\citeauthor{shirley02} and \citeauthor{young03} included contributions
to the heating of the envelope by the external interstellar radiation
field and adopted outer envelope radii significantly larger than those
set by the 10~K boundary used in \citetalias{jorgensen02}. For the
sources common to the two samples, \citeauthor{shirley02} found on
average steeper density profiles than ours for the class 0 objects
whereas \citeauthor{young03} found similar density profiles to ours
for the class I objects. \citeauthor{young03} suggested that the
disagreement for the class 0 objects and agreement for the class I
objects was due to a combination of neglect of the ISRF and an
underestimate of the sizes of the envelopes in
\citetalias{jorgensen02}: while inclusion of the ISRF will indeed tend
to steepen the derived density profile, an overestimate of the outer
radius (by factors of 2 or more) will tend to flatten the derived
profile. As illustrated above, however, these parameters have
negligible impact on the \object{IRAS2} envelope structure. It is
therefore interesting to note the agreement in slope between the
interferometer and SCUBA continuum observations, in contrast to the
discussion of \object{B335} by \cite{harvey03}. Comparing to the
results of \cite{shirley02}, \citeauthor{harvey03} found a slightly
flatter density profile when modeling the interferometer
observations. While uncertainty in the outer radius and ISRF may lead
to only small departures for the interferometry data, it can lead to
systematic changes in the slope of derived power-law density profile
from the SCUBA observations of $\simeq 0.2$. This could explain the
differences between the density profiles from the interferometry and
SCUBA data for \object{B335}.

\begin{figure}
\resizebox{\hsize}{!}{\includegraphics{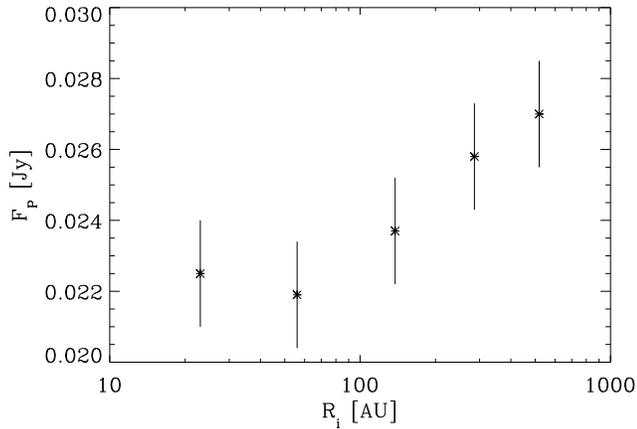}}
\caption{Derived point source flux plotted against size of the inner
envelope cavity.}\label{innerradius_pointsource}
\end{figure}

Our envelope model is entirely based on SCUBA data, and the
interferometer fluxes serve only to constrain any point source
flux. The robustness of that constraint depends on the assumption that
the envelope model can be extrapolated down to scales much smaller
than the SCUBA resolution (4\arcsec = 900~AU). In
\citetalias{jorgensen02} the inner radius is fixed at a temperature of
250~K which occurs at $R=22$~AU, but it was argued that this is not
determined by the data. The size of any inner cavity is expected to
affect the interferometer data since these sample small scales: a
larger adopted cavity would result in a higher inferred point source
flux to compensate for the reduced small-scale
emission. Fig.~\ref{innerradius_pointsource} plots this `required'
point source flux against inner cavity size. The point source flux
increases from 22~mJy for cavities $<25$~AU to $\approx 27$~mJy for
cavities $\gtrsim 200$~AU (1\arcsec). The interferometer data would
resolve cavities larger than this and a point source could no longer
compensate for the removed emission.

Turning this reasoning around, the inferred point source could be due
to an increase in envelope density on small scales, as opposed to a
circumstellar disk in an envelope cavity. Assuming a temperature of
150~K appropriate for the envelope scales unresolved by the
interferometers instead of the 30~K assumed for the disk, one derives
a mass of 0.06~$M_\odot$. For comparison the mass of the single
power-law density model within 150~AU is only 0.008~$M_\odot$. So to
explain the detected flux an increase in density by almost an order of
magnitude is needed, which seems unlikely.

\begin{figure}
\resizebox{\hsize}{!}{\includegraphics{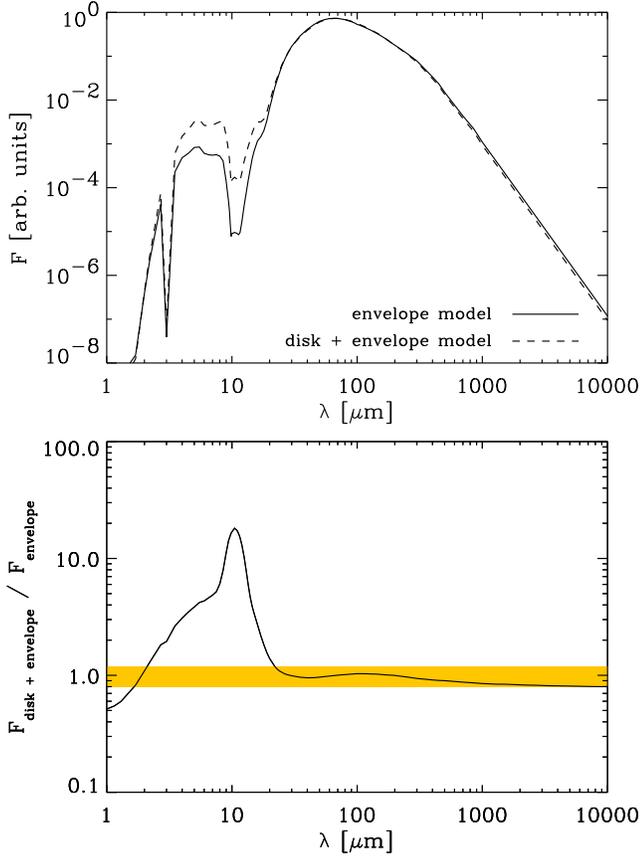}}
\caption{Changes of the emerging SED due to inclusion of a 200~AU
outer, 0.3 $M_\odot$ disk: in the upper panel the SEDs from the 
star + envelope (solid line) and star + envelope + disk models
(dashed line) are compared. In both models the central star is
represented by a 5000~K blackbody. In the lower panel the ratio
between the models are shown - the typical 20\% error-level in the
flux calibration is illustrated by the solid
rectangle.}\label{disk_sed}
\end{figure}

The model from \citetalias{jorgensen02} assumes that the envelope is
heated by a stellar blackbody of 5000~K at the center. If, as is
argued above, the star is surrounded by a disk that reprocesses a
significant fraction of the stellar light, the input spectrum shifts
to longer wavelengths. To investigate the effect on the envelope's
temperature structure, Fig.~\ref{disk_sed} compares the SEDs of the
original model and a model where the central star is surrounded by a
200~AU outer radius, 0.33~$M_\odot$ disk. The disk follows the
descriptions of \cite{chiang97} and \cite{dullemond01}, and the
envelope's inner cavity has been increased to 200~AU in radius so that
it encompasses the disk. As a result, the temperature at the inner
edge of the envelope drops from the original 250~K (at 22 AU) to 75~K
(at 200~AU). The radiative transfer code DUSTY produces the envelope's
temperature distribution and emergent SED using the star+disk spectrum
as heating input, similar to the calculations of
\citetalias{jorgensen02} for the star-only spectrum. The comparison in
Fig.~\ref{disk_sed} shows that the SEDs between 60~$\mu$m and 1.3~mm
are unchanged. Our derived envelope parameters are therefore
unaffected by the exact form of the input spectrum. The departures
grow larger at the shorter wavelengths (2--20~$\mu$m) and may be
observable with, e.g., SIRTF. It is not surprising that the SEDs are
most different at these wavelengths. Flared disk models such as those
of \cite{chiang97} are specifically invoked to explain so-called
`flat-spectrum' sources. Their superheated surface layers `flatten'
the SED of these star+disk systems by boosting the 2--20~$\mu$m
emission. It is not obvious that such a description of the disk
is valid for early, deeply embedded objects, such as IRAS2A. Still,
the important point here is that the influence of the disk on the
observed SED is likely to be negligible at the wavelengths where the
envelope model is constrained.

\subsection{A collapse model for the continuum emission}\label{shusolution}
As demonstrated by \cite{hogerheijde00sandell}, \cite{shirley02}, and
\cite{schoeier02} models other than a density power law can also fit
continuum observations, in particular the inside-out collapse model of
\cite{shu77}. These authors conclude that a collapse model can provide
an equally good fit as power-law models, with the caveat by
\cite{shirley02} that collapse models only fit their class~0 objects
for sufficiently low ages where this model is well approximated by a
single power law on the scales resolved by SCUBA. \cite{schoeier02}
and \cite{shirley02} find that continuum and line data sets sometimes
give discrepant collapse model fits to the same sources, with line
data favoring higher ages than continuum data.

\begin{figure}
\resizebox{\hsize}{!}{\includegraphics{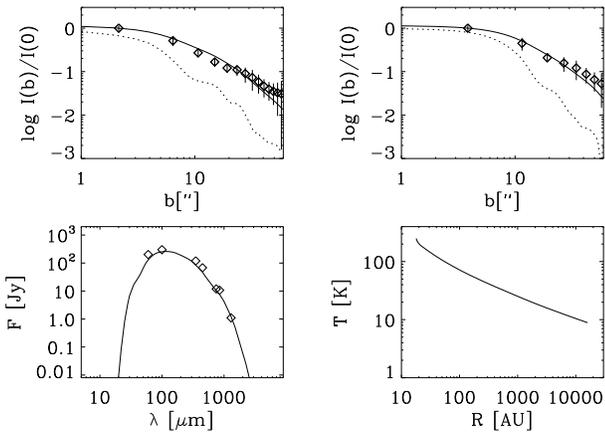}}
\caption{Fits to the SCUBA observations with a inside-out collapse
model with an isothermal sound speed, $a$, of 0.3~\kms\ and an age of
$1.7\times 10^4$~years.}\label{shumodel}
\end{figure}

Fitting the \cite{shu77} inside-out collapse model to the SCUBA data
for \object{IRAS2} gives best fit values of $a=0.3$ \kms\ and
$t=1.7\times 10^{4}$ years (see Fig.~\ref{shumodel}) with the quality
of the fits essentially identical to those of the single power-law
models. This collapse model also fits the BIMA and OVRO data if a
point source of 25~mJy is introduced.

\begin{figure}
\resizebox{\hsize}{!}{\rotatebox{-90}{\includegraphics{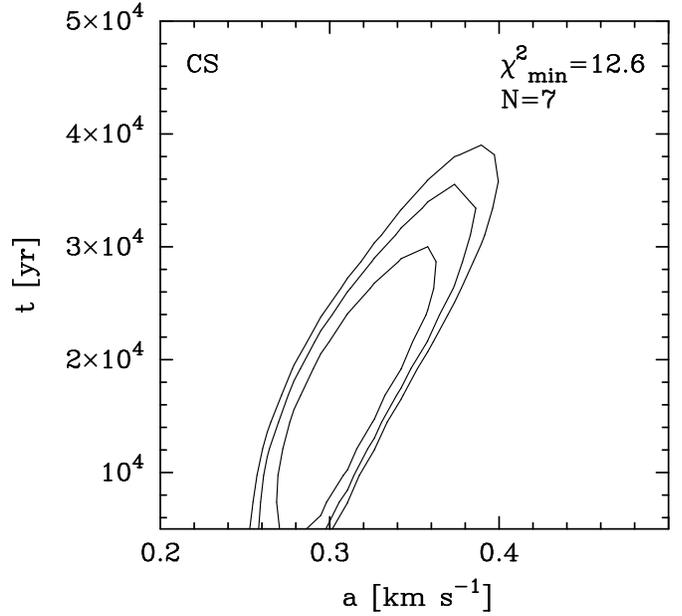}}}
\caption{Constraints on the inside-out collapse model derived from the
CS and \cfs\ line intensities, assuming CS and \cfs\ abundances of
1\tpt{-9} and 1\tpt{-10} respectively found in the single power-law
density model. The inferred age and sound speed agree well with those
derived from the continuum data.}\label{shumodel_cs}
\end{figure}

The integrated CS and \cfs\ line intensities were fitted independently
with the collapse model using detailed radiative transfer as in
\cite{schoeier02} for a constant fractional abundance with radius. The
uncertainties in the line fluxes are assumed to be 20\% and for each
model the $\chi^2$-estimator is used to pick out the best model and
estimate confidence levels for the derived parameters. Interestingly,
the fits to the CS and \cfs\ lines (Fig.~\ref{shumodel_cs}) give
identical parameters to those derived from the dust modeling, in
contrast with the other sources \citep[e.g.][]{shirley02,schoeier02}.

The identical fits to the line intensities and continuum observations
and success of both collapse and power-law density models illustrates
the low age inferred for \object{IRAS2}: the collapse expansion radius
in the inside-out collapse model is for the envelope parameters
located at $\approx 4\arcsec$ and therefore not directly probed by the
SCUBA continuum maps. Likewise the CS lines predominantly probe the
outer regions of the envelopes where the density profile in the
collapse model for the low age of \object{IRAS2} essentially is a
power-law. This also explains why a slightly higher point source flux
is obtained with the collapse model than the power-law density model:
inside the collapse expansion radius the density profile flattens in
the collapse model, lowering the mass and thereby the flux towards the
unresolved center of the envelope. This is compensated by increasing
the point source flux when modeling the interferometer observations.

In summary, the interferometer continuum data are well described by
the same 1.7~$M_\odot$ envelope models that fit the SCUBA
data. Power-law descriptions for the density and inside-out collapse
model fit the data equally well. They indicate the presence of a
$F_{\rm 3\,mm}\approx$~22~mJy point source, presumably a
$\gtrsim$~0.33~$M_\odot$ circumstellar disk. Uncertainties associated
with the envelope model are reflected in the accuracy of the point
source flux, which may vary by up to 25\% from the quoted value. The
next section describes the line emission in this context.

\section{Line emission}\label{lines}
\subsection{Morphology}
\begin{figure*}
\resizebox{\hsize}{!}{\includegraphics{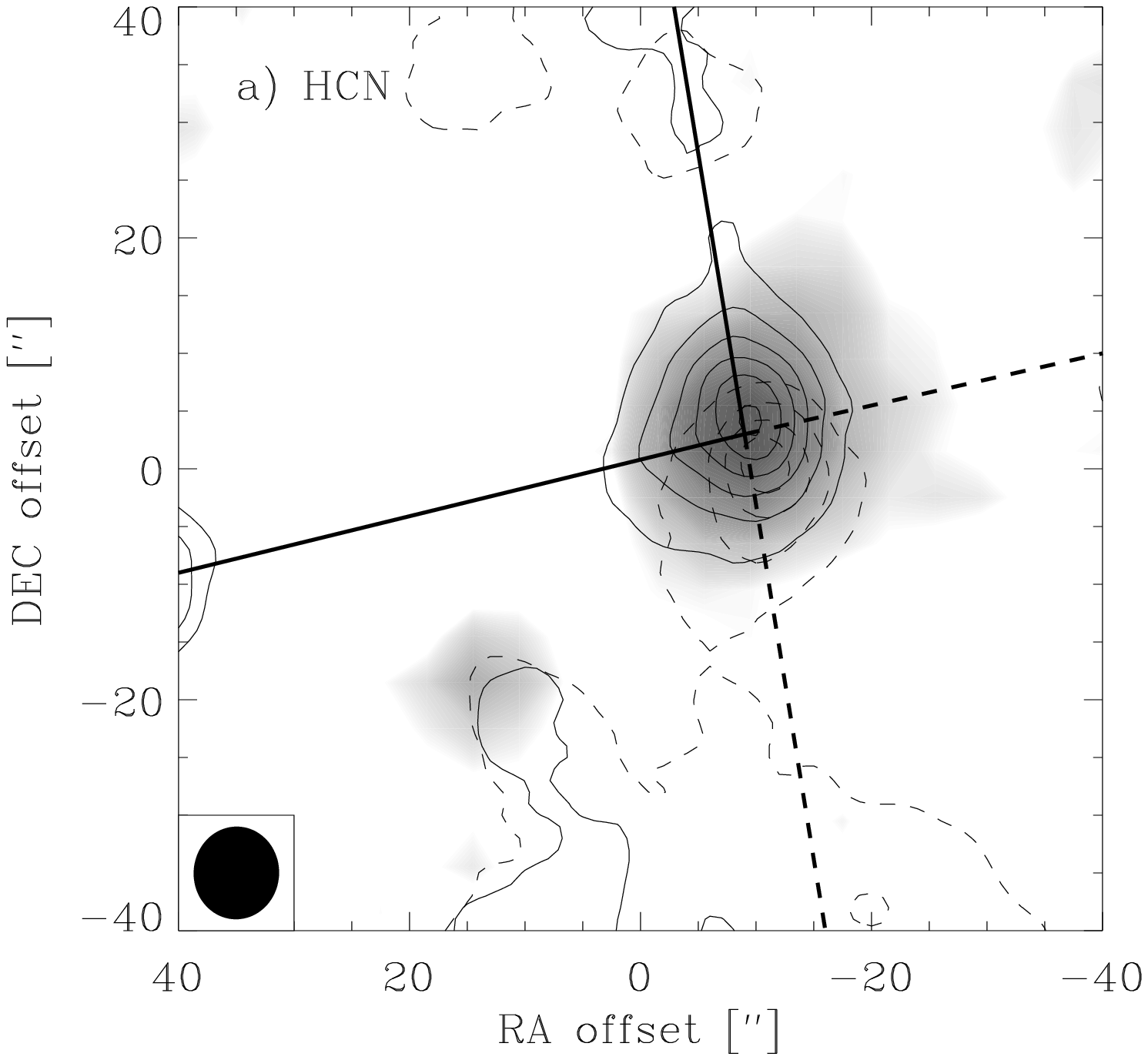}\includegraphics{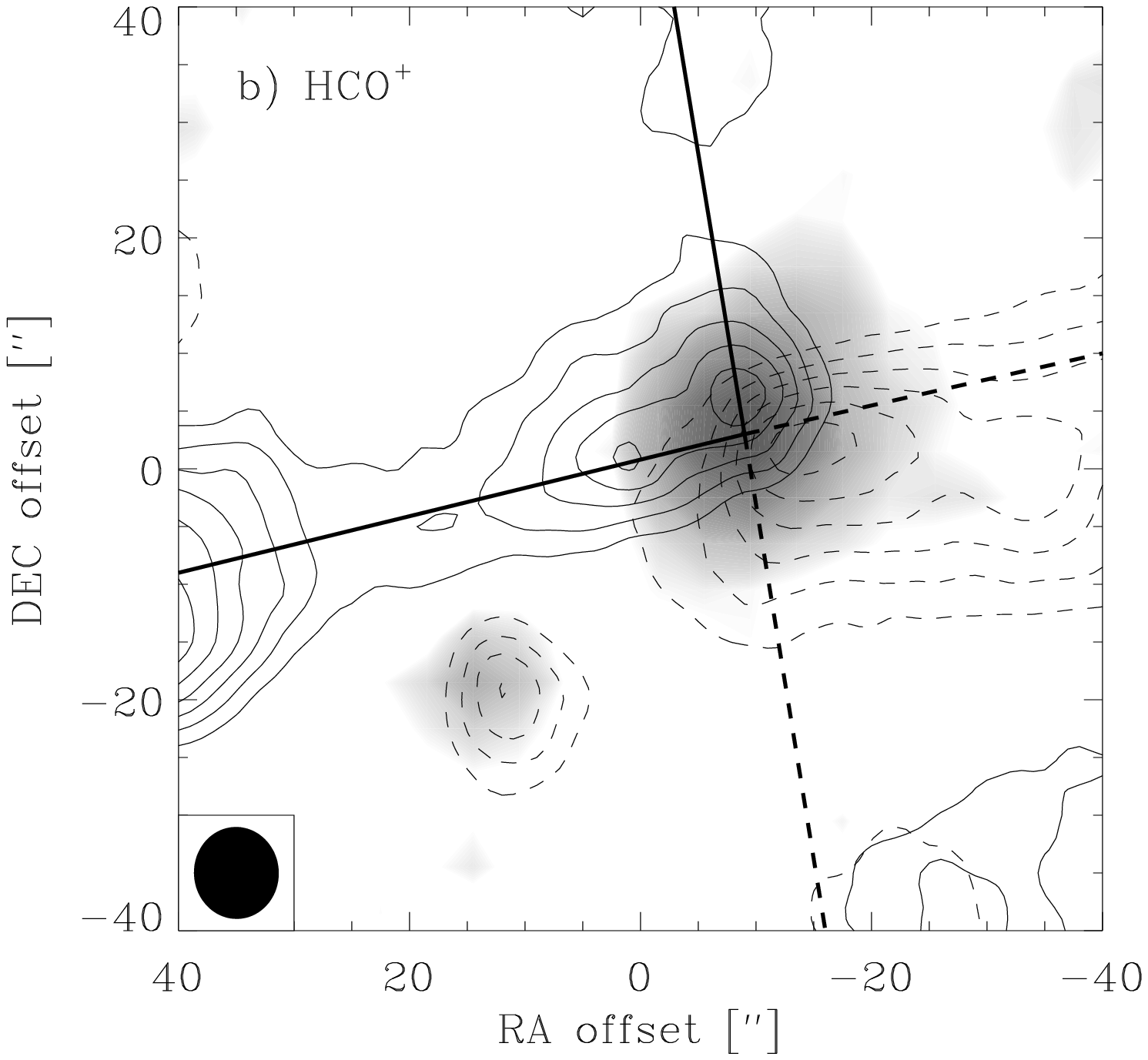}}
\resizebox{\hsize}{!}{\includegraphics{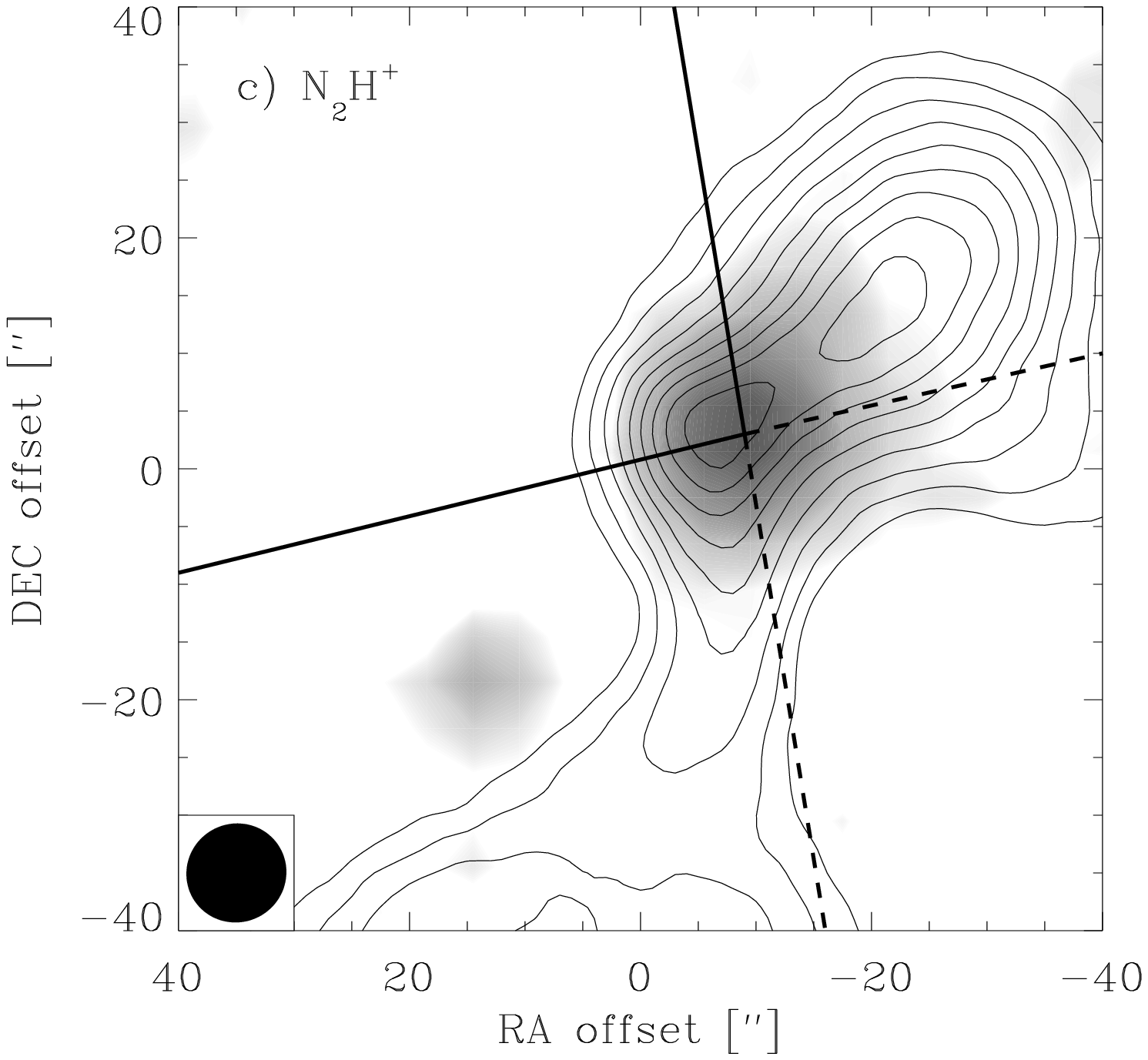}\includegraphics{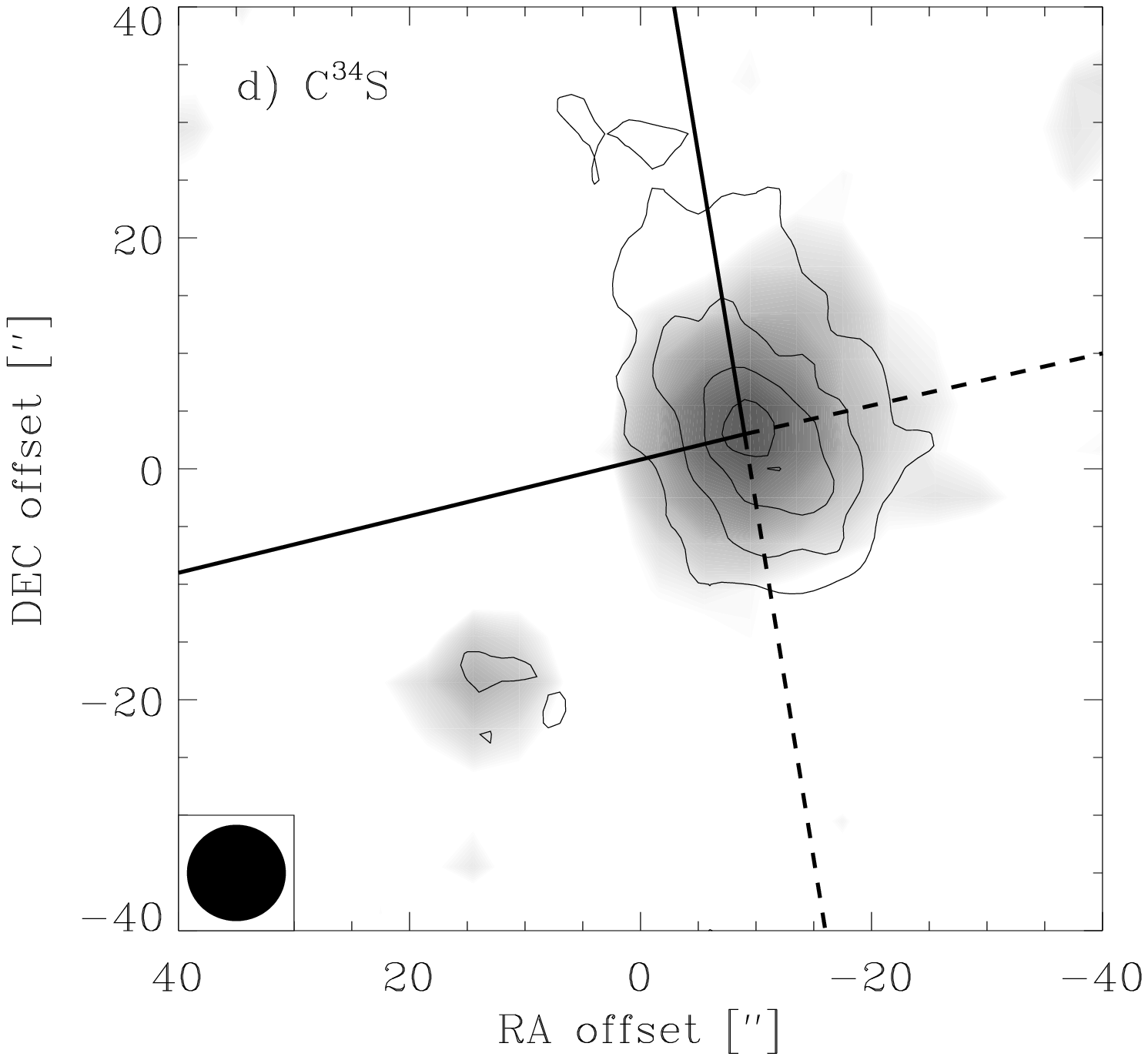}}
\caption{Integrated line emission from the BIMA observations for a)
HCN, b) \hcop, c) \nthp and d) \cfs\ plotted over the 3~mm continuum
maps (grey-scale). The outflow axes have been marked with straight
lines with the red part being solid and blue part being dashed. For
HCN and \hcop\ the emission has been integrated over the red and blue
parts of the line (3 to 7~\kms\ and 9 to 13~\kms) shown as the dashed
and solid lines, respectively. For \nthp\ and \cfs\ the total
integrated emission is presented, in the case of \nthp\ integrated
over the main group of hyperfine lines. For \cfs\ the contours are
presented in steps of 3$\sigma$, for HCN and \hcop\ in steps of
5$\sigma$ and for \nthp\ in steps of 7$\sigma$. }\label{bima_moment}
\end{figure*}

\begin{figure*}
\resizebox{\hsize}{!}{\includegraphics{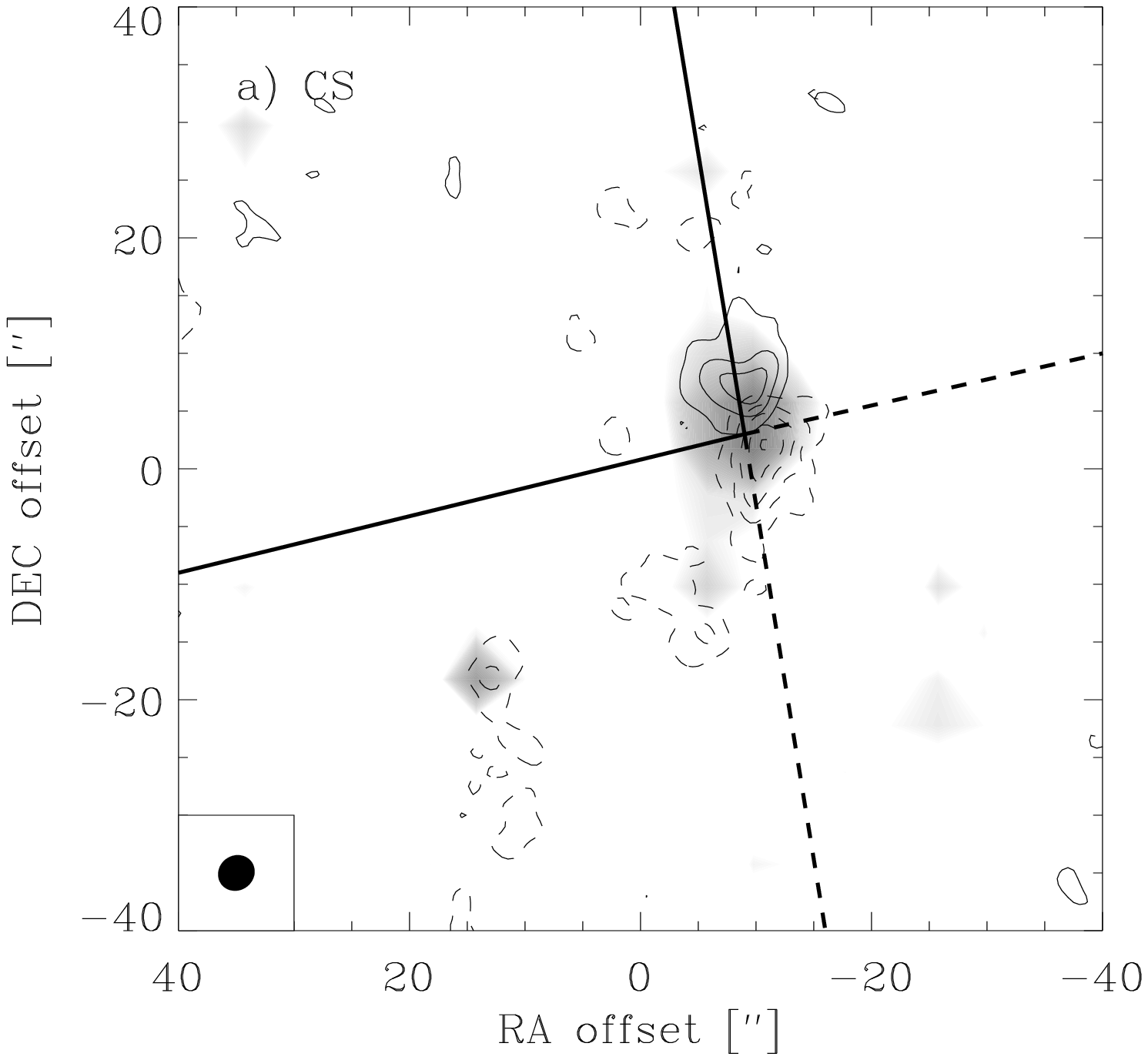}\includegraphics{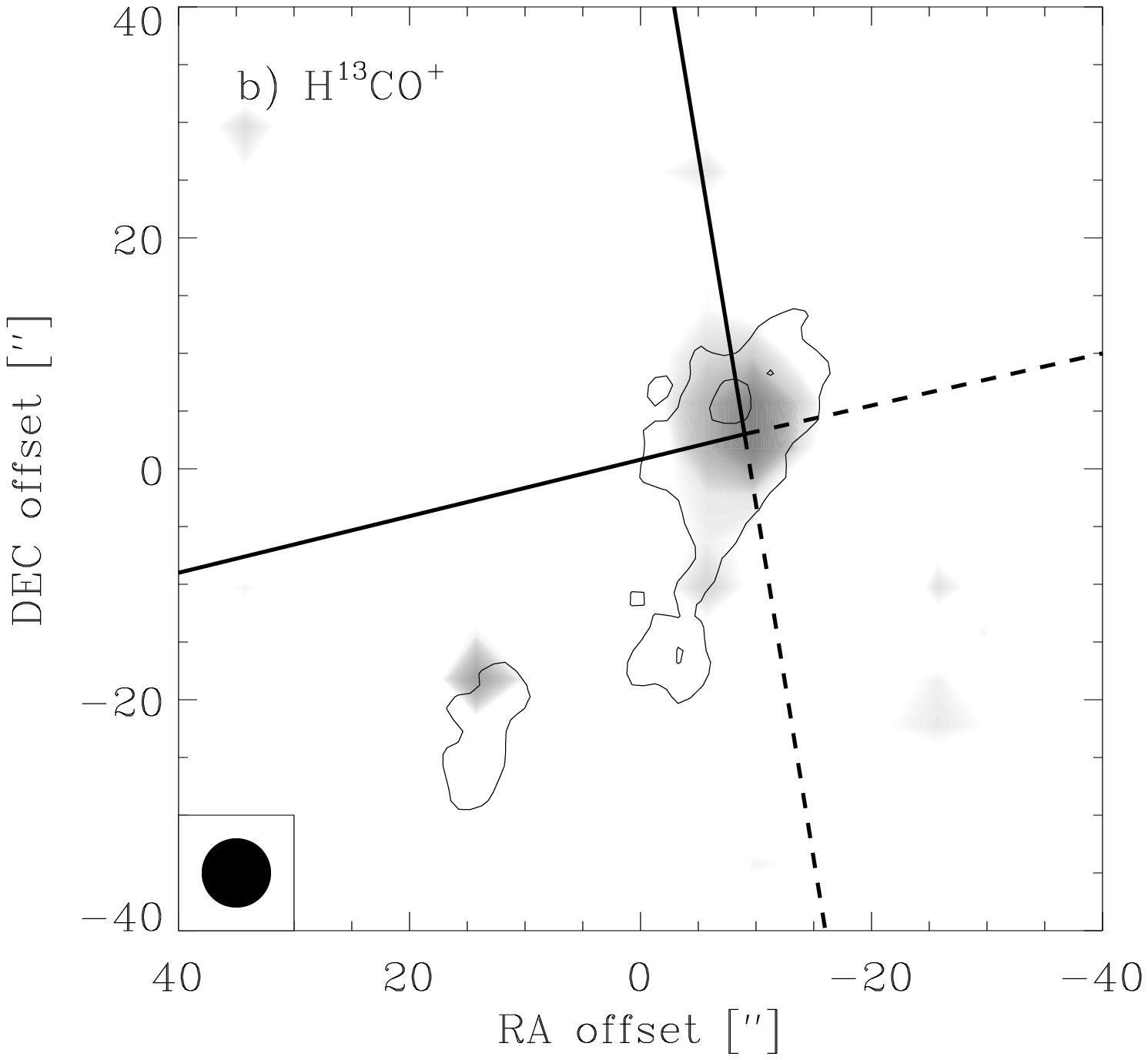}}
\resizebox{\hsize}{!}{\includegraphics{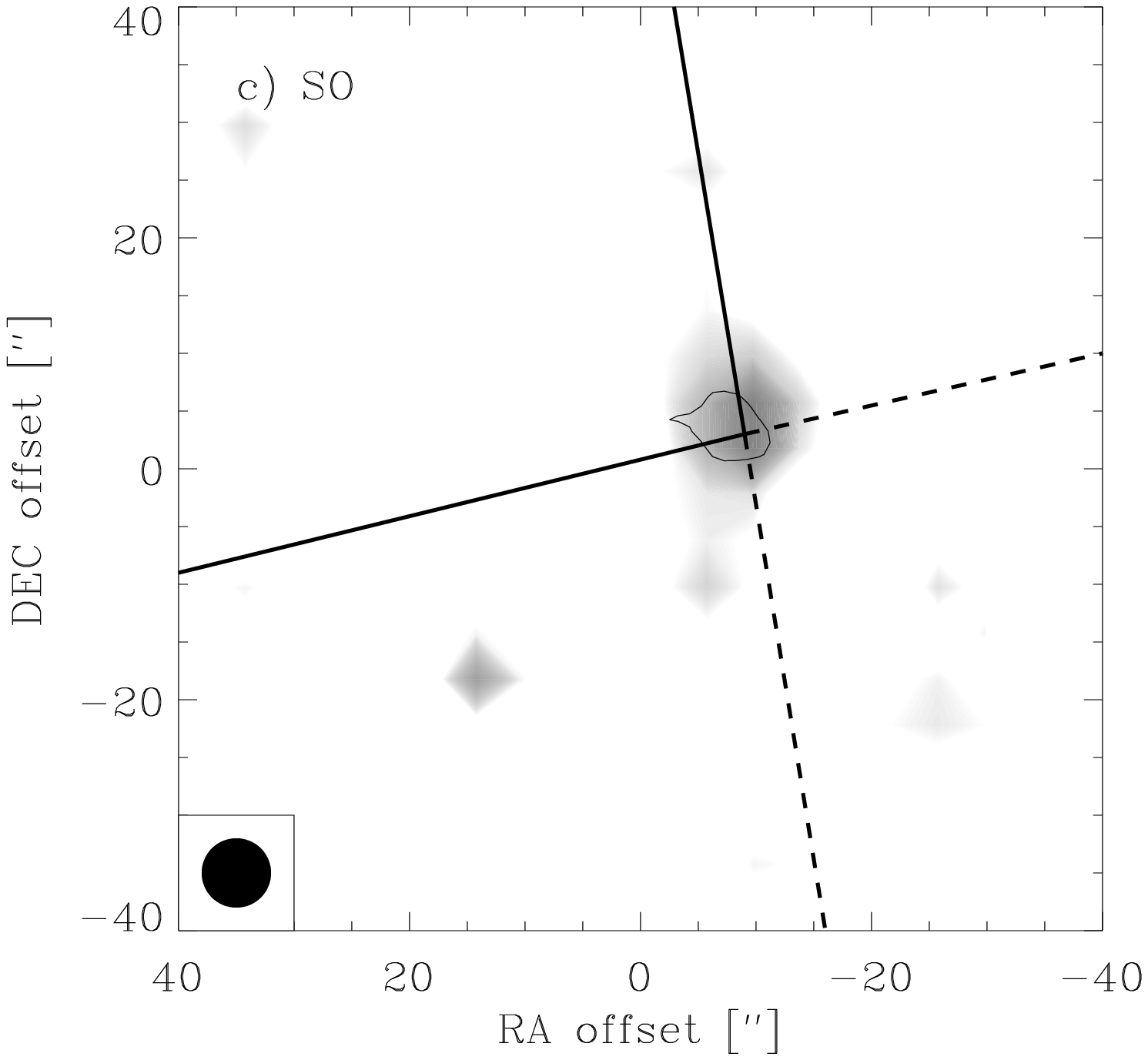}\includegraphics{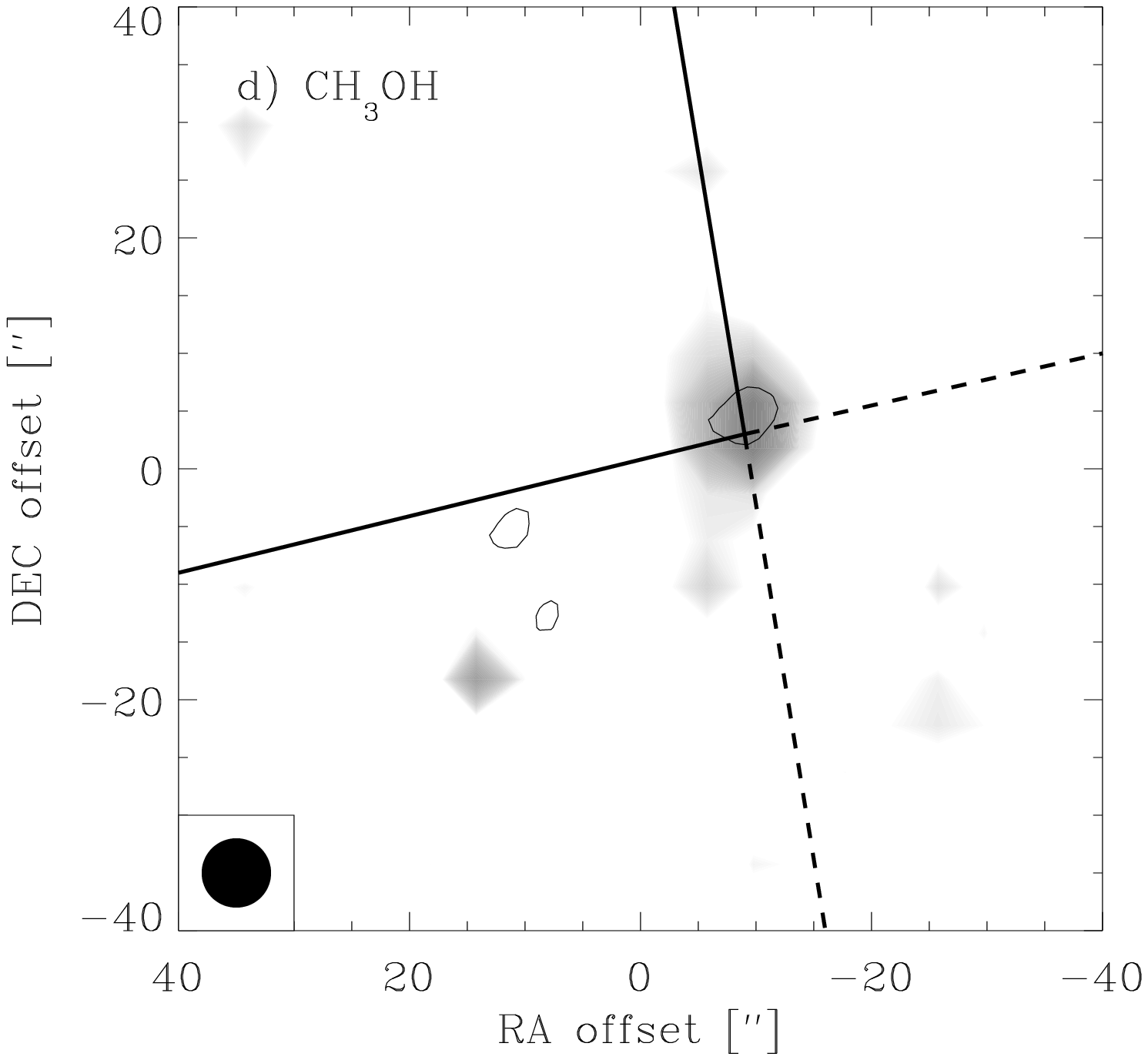}}
\caption{Integrated line emission from the OVRO data, showing a) CS,
b) \htcop, c) SO and d) \chtoh. CS is integrated over blue (5 to
9~\kms; dashed contours) and red (9 to 13~\kms; solid contours) parts
of the line with contours in steps of 3$\sigma$. For the other
molecules lines have been integrated from 7 to 11~\kms\ and contours
are given in steps of 2$\sigma$. The grey-scale indicate the 3~mm
continuum maps.}\label{ovro_moment}
\end{figure*}

\begin{figure*}
\resizebox{\hsize}{!}{\includegraphics{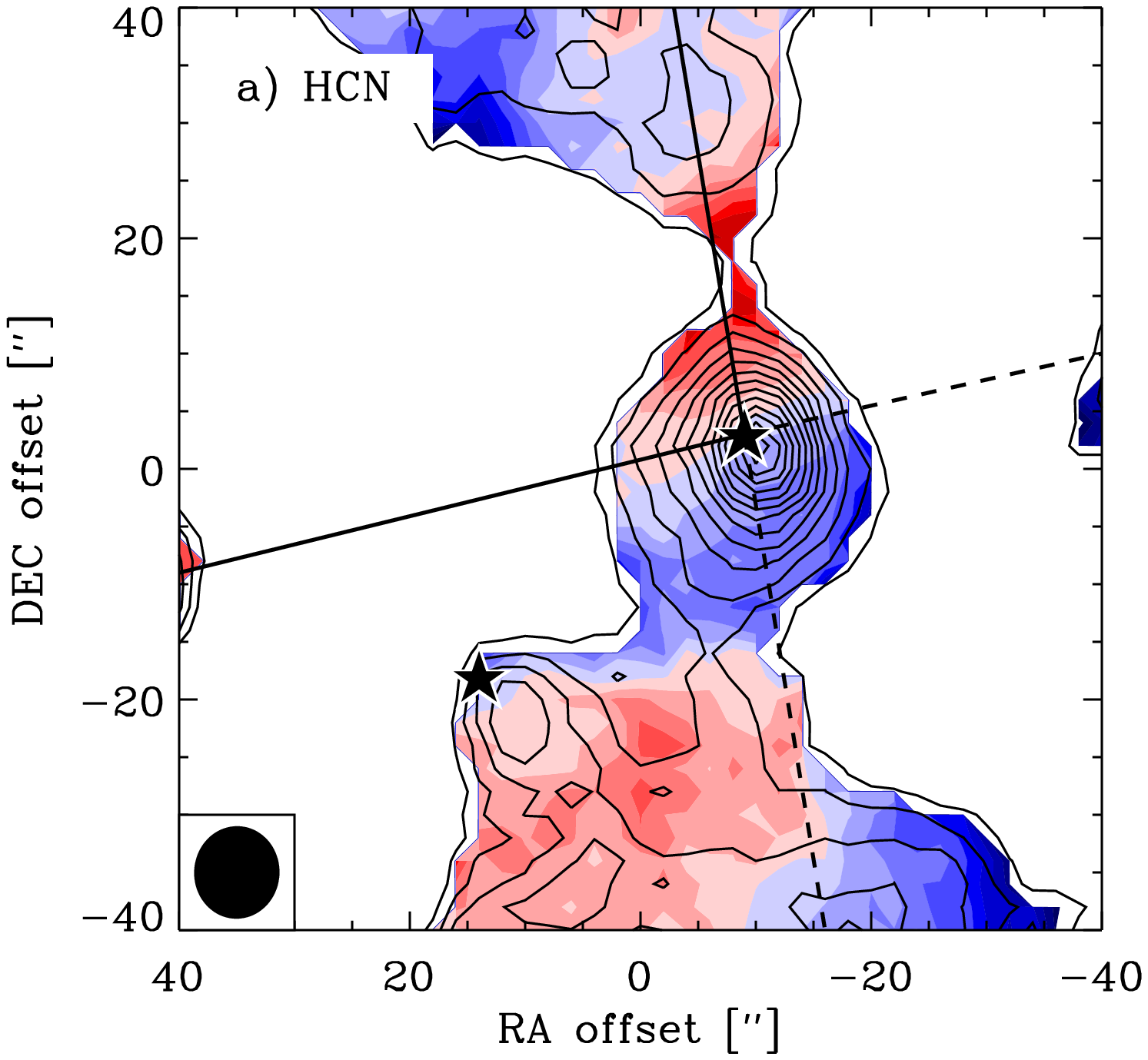}\includegraphics{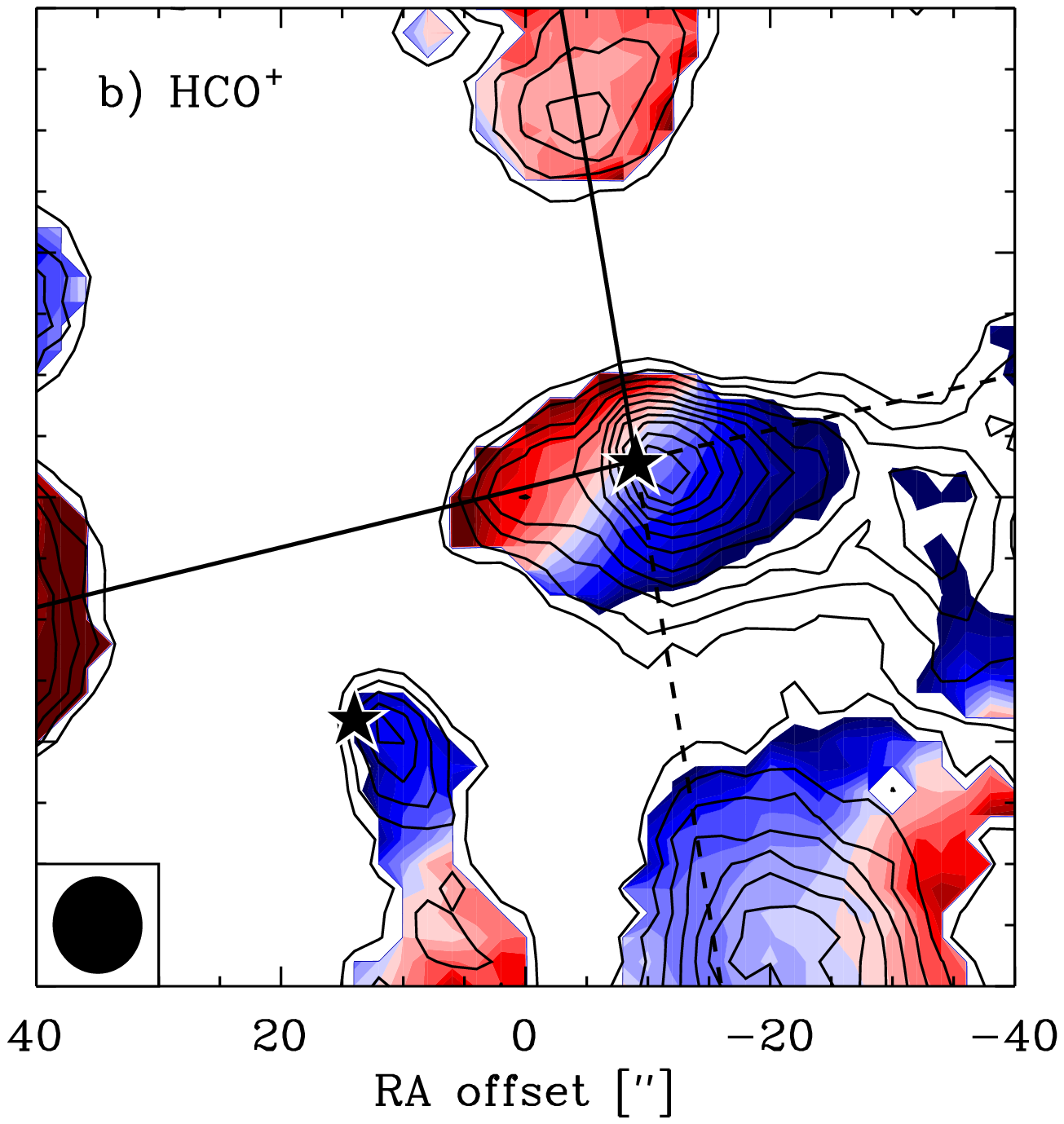}\includegraphics{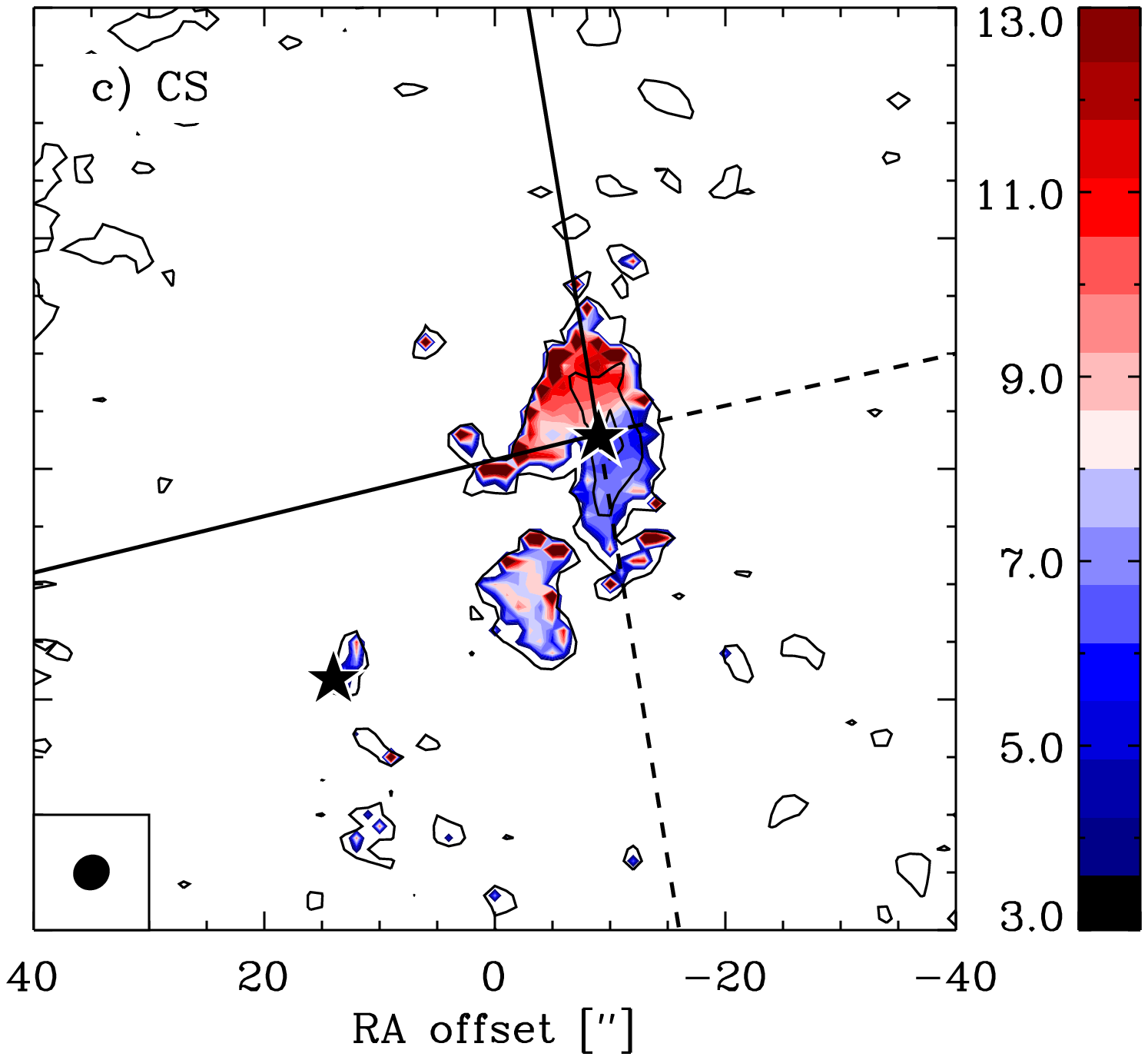}}
\caption{First order moment (velocity) maps of the a) HCN, b) HCO$^+$
(BIMA) and c) CS emission (OVRO). Each map has been overplotted with
the total integrated emission in steps of 3~$\sigma$ (solid line
contours). The outflow axes have been marked by lines and
\object{IRAS2A} and \object{IRAS2B} continuum sources by
stars.}\label{first_order}
\end{figure*}

Fig.~\ref{bima_moment}-\ref{ovro_moment} shows the integrated
intensity maps of all lines detected with BIMA (HCN, HCO$^+$,
N$_2$H$^+$, C$^{34}$S) and OVRO (CS, H$^{13}$CO$^+$, SO, and
CH$_3$OH). Velocity centroid images are shown for CS, HCN and HCO$^+$
in Fig.~\ref{first_order}. Emission of SiO and SO$_2$ was not detected
toward the source position. The images from BIMA show more extended
structure than those from OVRO because of the different $(u,v)$
coverage of the two arrays. All detected lines have a peak near the
object 2A, and most show a peak near 2B. The non-detections of SO and
CH$_3$OH near 2B are likely due to limited sensitivity given the low
signal-to-noise of these lines toward 2A. Only the non-detection of
N$_2$H$^+$ toward 2B is highly significant: emission in this line
appears to avoid 2B. The extended emission picked up by BIMA shows
three components. A roughly north-south ridge seen in HCN, HCO$^+$,
and N$_2$H$^+$; emission along the east-west outflow in HCO$^+$ and,
at the tip of the eastern jet at the edge of the image, in HCN; and an
extended peak in N$_2$H$^+$ around the continuum position
2C. Interestingly, N$_2$H$^+$ also seems to avoid the east-west
outflow and appears almost anti-correlated with HCO$^+$.

\begin{figure}
\resizebox{\hsize}{!}{\includegraphics{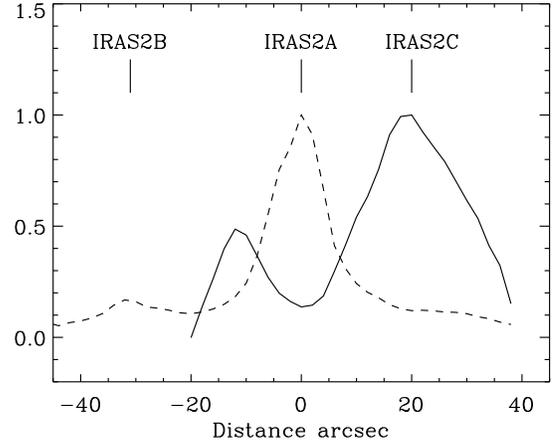}}
\caption{The contrast between the \nthp\ and SCUBA emission: the \nthp
emission divided by the 450~$\mu$m continuum emission (normalized)
along a straight line with a position angle of 45$^\circ$ through
\object{IRAS2A} (solid line) and the 450~$\mu$m continuum emission in the same
positions (dashed line).}\label{nthpcontrast}
\end{figure}

The intensity ratio of 1.5:2.7:4.7 of the N$_2$H$^+$ hyperfine lines
suggest that the emission is optically thin and close to LTE, where
the ratio would be 1:3:5. Relative to the 450~$\mu$m emission from
SCUBA that traces cool dust, the N$_2$H$^+$ emission is strongest
around 2C, lower around 2A, and absent toward 2B as illustrated in
Fig.~\ref{nthpcontrast}. In a study of dark cloud cores,
\cite{bergin01,bergin02} find that N$_2$H$^+$ has a large abundance
deep inside the clouds and a lower abundance in the exterior
regions. This trend is opposite to that of CO, which is often highly
depleted deep inside cores. \citeauthor{bergin01} argue that
N$_2$H$^+$ is effectively destroyed through reactions with CO, and
therefore is only present at high abundance where CO is depleted. This
scenario can also explain the relative distribution of N$_2$H$^+$ in
2C, 2A, and 2B: in the starless core 2C temperatures are low and CO is
highly depleted, resulting in strong N$_2$H$^+$ emission; in 2A the
star has already heated the material and some CO has been released,
reducing the N$_2$H$^+$ abundance and emission; the evolved core 2B
has been thoroughly heated by the star, and most N$_2$H$^+$ has been
effectively destroyed by the released CO. This hints at triggered star
formation with the sources lining up from southeast to northwest in
evolutionary order, with 2B older than 2A, and 2A older than 2C.

\begin{figure}
\resizebox{\hsize}{!}{\includegraphics{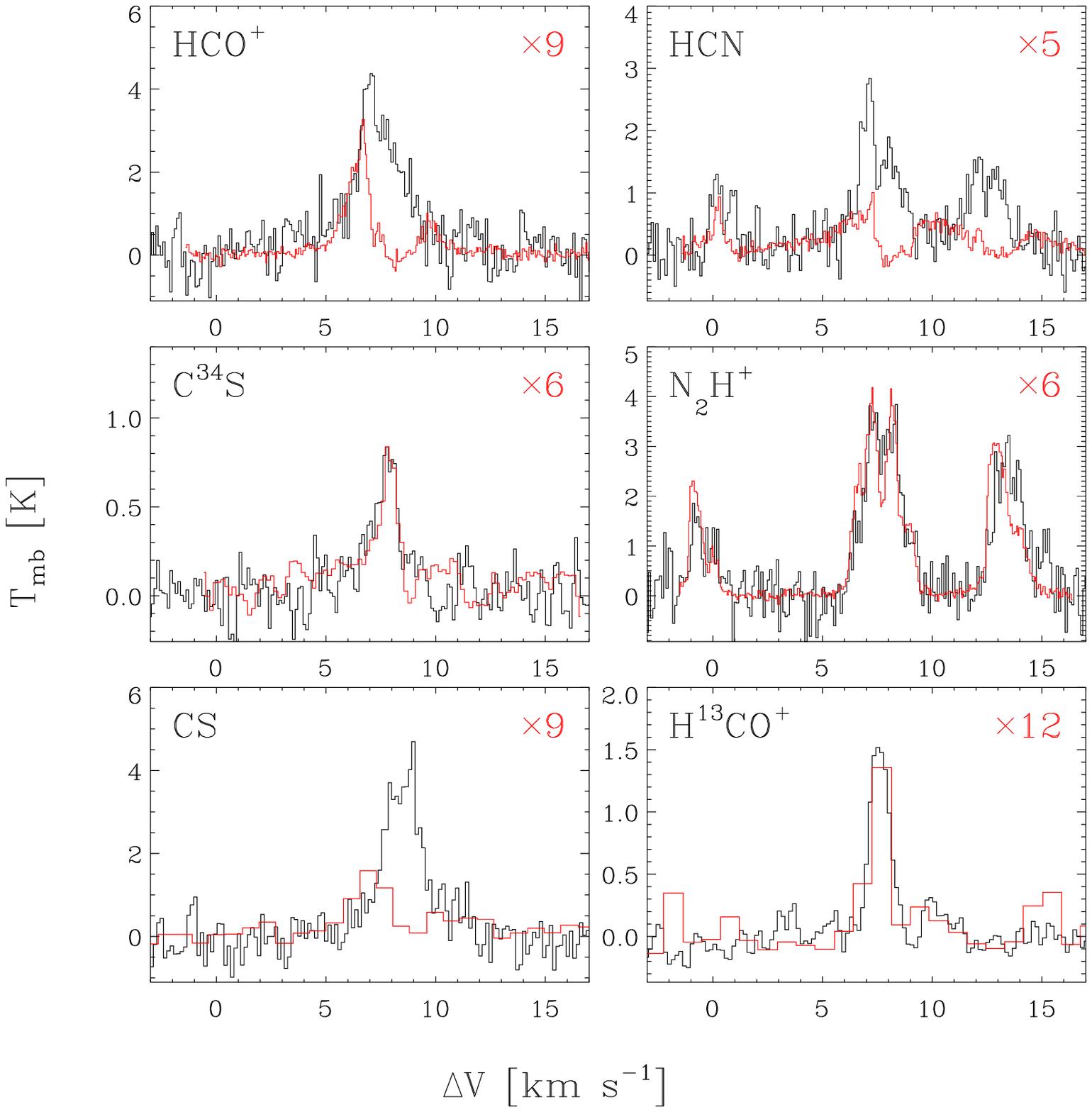}}
\caption{Comparison between the single-dish observations (dark) and
corresponding spectra from the interferometer observations restored
with the single-dish beam (red). The spectra from the interferometry
observations have been scaled by the factors indicated in the upper
right corner (factors 5--12) to include all spectra in the same
plots.}\label{resolved_out}
\end{figure}

Several molecules only show emission in a very narrow velocity range
around the systemic velocity of the cloud of $\sim 8$ km~s$^{-1}$:
H$^{13}$CO$^+$, SO, CH$_3$OH, N$_2$H$^+$, and C$^{34}$S. Others show
pronounced gradients in a north-south direction within
$\approx$~20\arcsec\ from IRAS2A (HCN, HCO$^+$, CS) and along the
east-west outflow (most clearly in HCO$^+$). Whether the north-south
gradient around 2A is rotation in a circumstellar envelope or is
related to the north-south outflow seen on larger scales is addressed
in Section~\ref{velgradient}. To asses the amount of recovered flux as
function of velocity, Fig.~\ref{resolved_out} compares single-dish
spectra with interferometer spectra averaged over the single-dish beam
size and converted to the antenna-temperature intensity scale. The
interferometer recovers at most 17\% of the emission, and much less in
many cases. Apart from a scaling factor, the line shapes of C$^{34}$S,
N$_2$H$^+$, and H$^{13}$CO$^+$ are similar in the interferometer
and single-dish spectra, implying that although the
interferometers picked up only the more compact emitting structures,
only small velocity gradients can be present within the envelope
itself. Deep self-absorption apparent in HCO$^+$, HCN, and CS near
the systemic velocity indicates that surrounding cloud material is
optically thick and entirely resolved out. The velocity structure seen
in these lines therefore only reflects material at relatively extreme
red- and blue-shifts.

\subsection{Envelope contributions to the line emission}
The available envelope model can help explain the interferometric line
data, by separating the expected emission of the envelope from other
components. We adopt the power-law envelope model of
Sect.~\ref{continuum}, and fix the molecular abundances from fitting
the integrated intensities of single-dish observations of optically
thin isotopes. Table~\ref{assumed_abundances} lists the derived
abundances. \cite{paperii} discuss this method in greater detail and
compare the results for a larger sample. With these abundances, the
molecular excitation is solved using the code of
\cite{hogerheijde00vandertak} and the emergent sky-brightness
distribution calculated. For comparison with the interferometer data
the sky-brightness distribution was sampled at the $(u,v)$ positions
of the data. The resulting visibilities were then inverted, cleaned,
and restored in a similar way as the data.
\begin{table}
\caption{Abundances in the 1D static envelope model derived from of
single-dish line observations. For details see
\cite{paperii}.}\label{assumed_abundances}
\begin{tabular}{ll} \hline\hline
Molecule & Abundance ([X/\hto]) \\ \hline
\cfs\                    & 1.4\tpt{-10} \\
CS                       & 1.3\tpt{-9}  \\
\htcop\ (all except 1-0 line) & 4.3\tpt{-11} \\
\htcop\ (1-0 line)       & 8.0\tpt{-11} \\ \hline
\end{tabular}
\end{table}

\begin{figure}
\resizebox{\hsize}{!}{\rotatebox{90}{\includegraphics{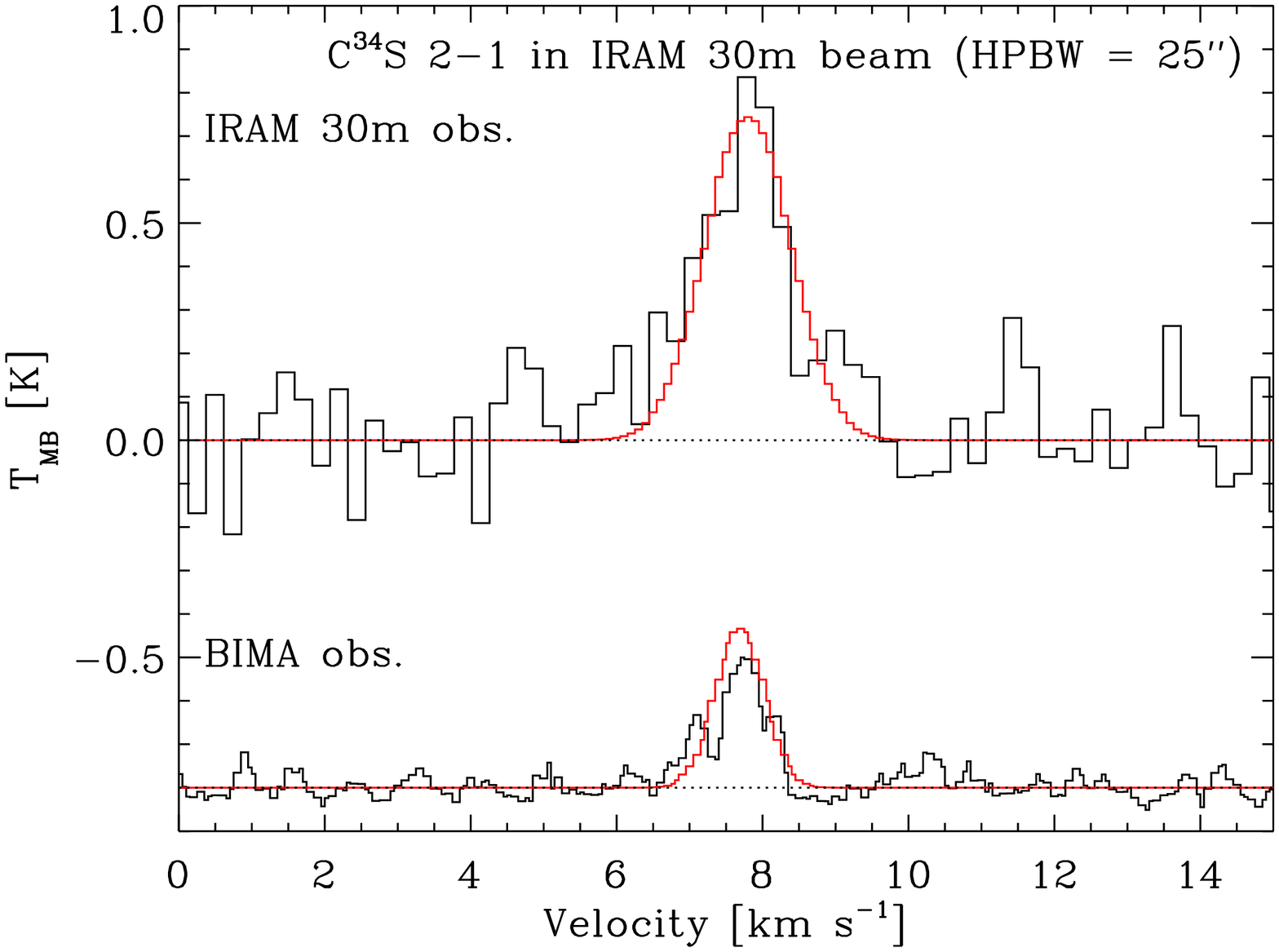}}}
\resizebox{\hsize}{!}{\includegraphics{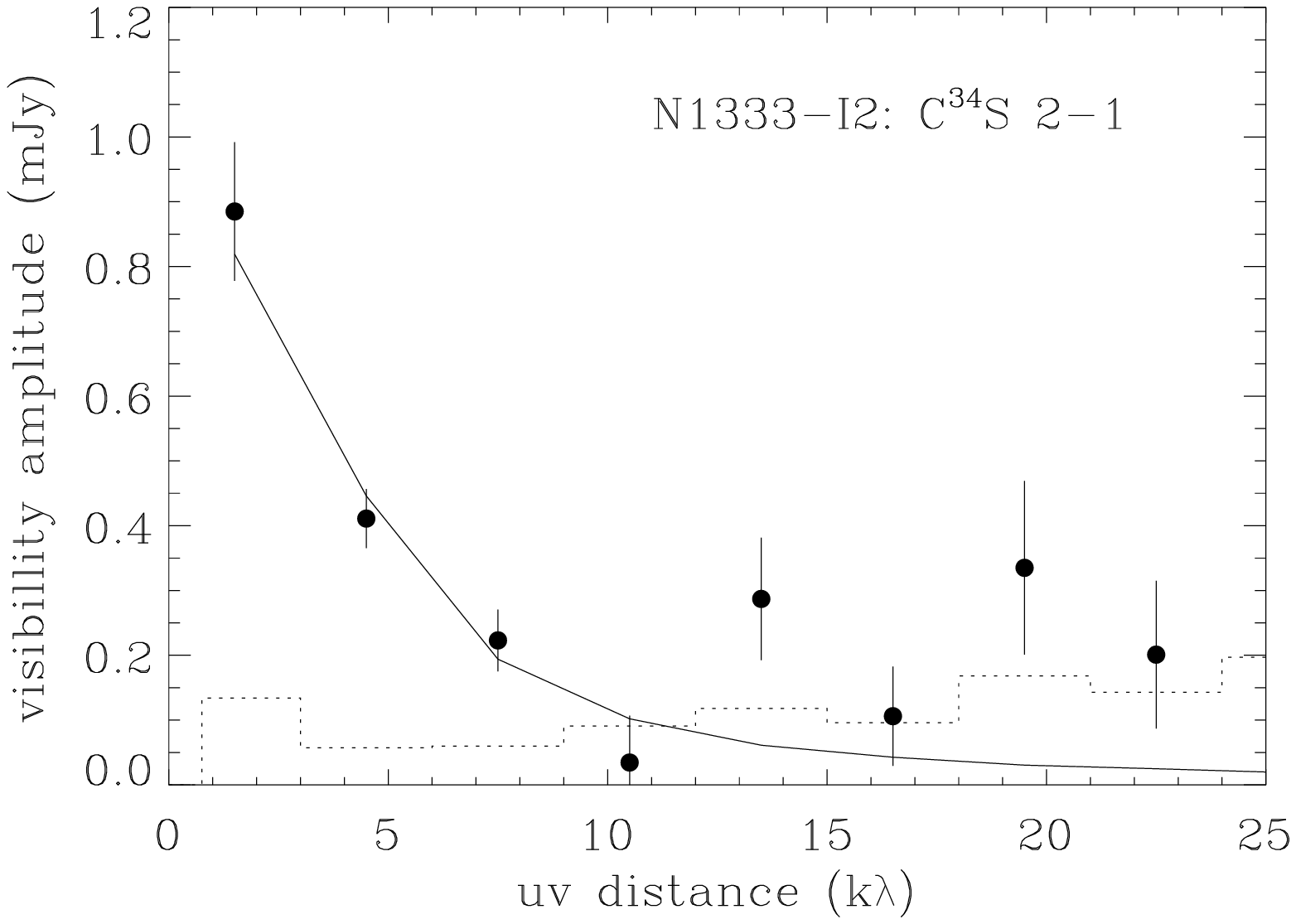}}
\caption{Upper panel: Comparison between the \cfs\ emission from the
single-dish observations using the IRAM 30~m telescope with the BIMA
interferometry observations (lower spectrum), offset along the $T_{\rm
mb}$ axis by -0.8~K and restored with a 25\arcsec\ beam similar to the
single-dish data. The prediction from the 1D static model of the
emission brightness distribution from \citetalias{jorgensen02} with
abundances derived from single-dish line observations given in
Table~\ref{assumed_abundances} and sampled at the relevant $(u,v)$
grid has been overplotted on the spectrum in red. Lower panel:
visibility amplitude plotted as function of projected baseline
length. The predictions from the envelope model with \cfs\ abundance
constrained by the single-dish line observations have been overplotted
as the solid line. The zero-expectation level is indicated as the
dotted histogram.}\label{cfs_interferometry}
\end{figure}

Fig.~\ref{cfs_interferometry} compares the observed and modeled line
emission for \cfs. The upper panel shows a comparison between the
model and the single-dish observations (upper spectrum) and the
interferometry data convolved with the single-dish beam (lower
spectrum). In the lower panel the visibilities are plotted as a
function of projected baseline length. Both comparisons show that the
model works very well in describing the interferometry and single-dish
line observations simultaneously and reproduces the emission
distribution at the observed scales. This has two implications. First,
that optically thin species such as \cfs\ trace material in the envelope
and are well described by the model derived from the continuum
observations. Second, for species such as \cfs\ the chemistry is
homogeneous at the observed radial scales in the envelope, so that a
constant fractional abundance is sufficient to describe the chemistry
within the assumptions and uncertainties in the modeling.

\begin{figure}
\resizebox{\hsize}{!}{\rotatebox{90}{\includegraphics{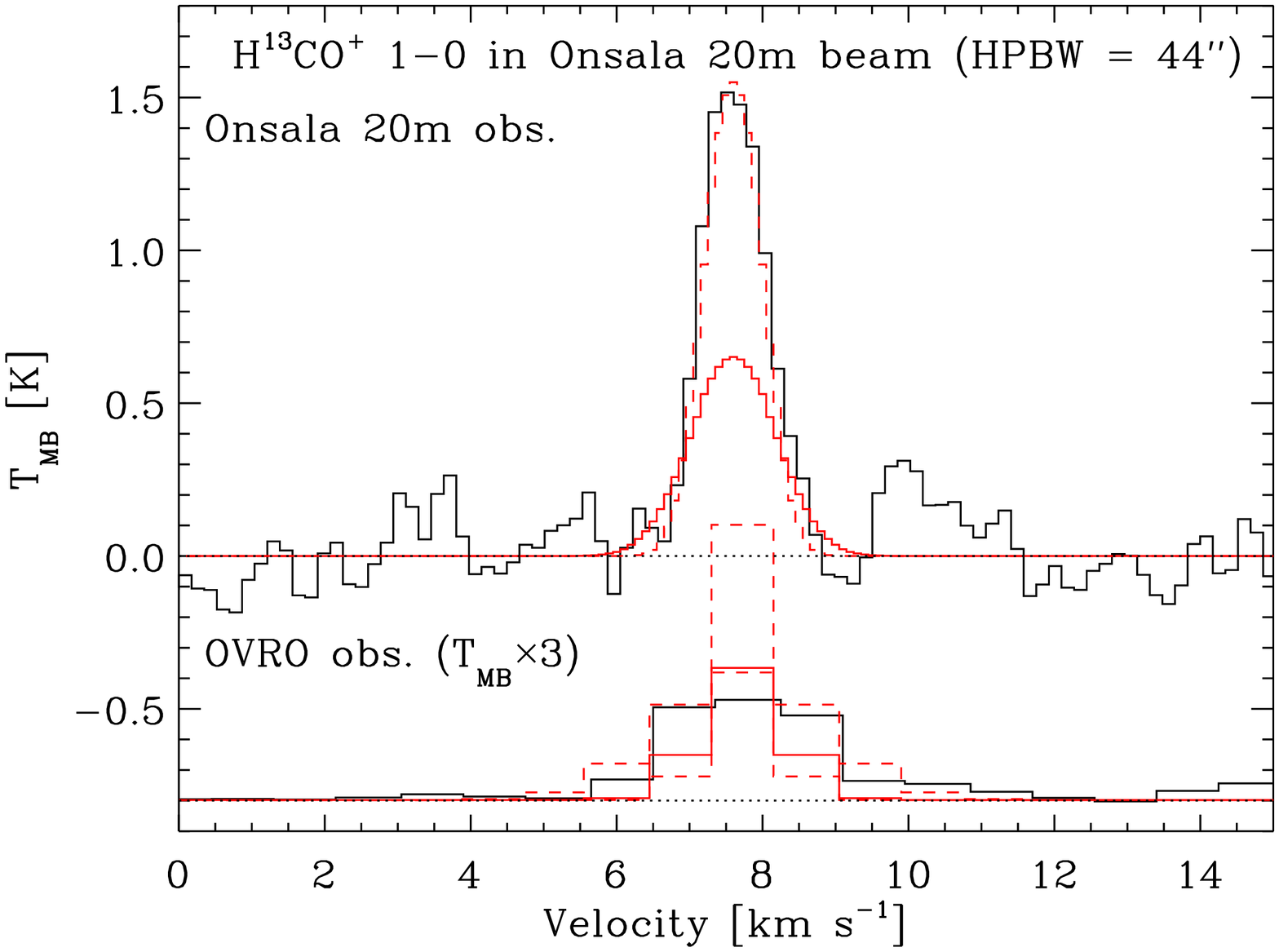}}}
\resizebox{\hsize}{!}{\includegraphics{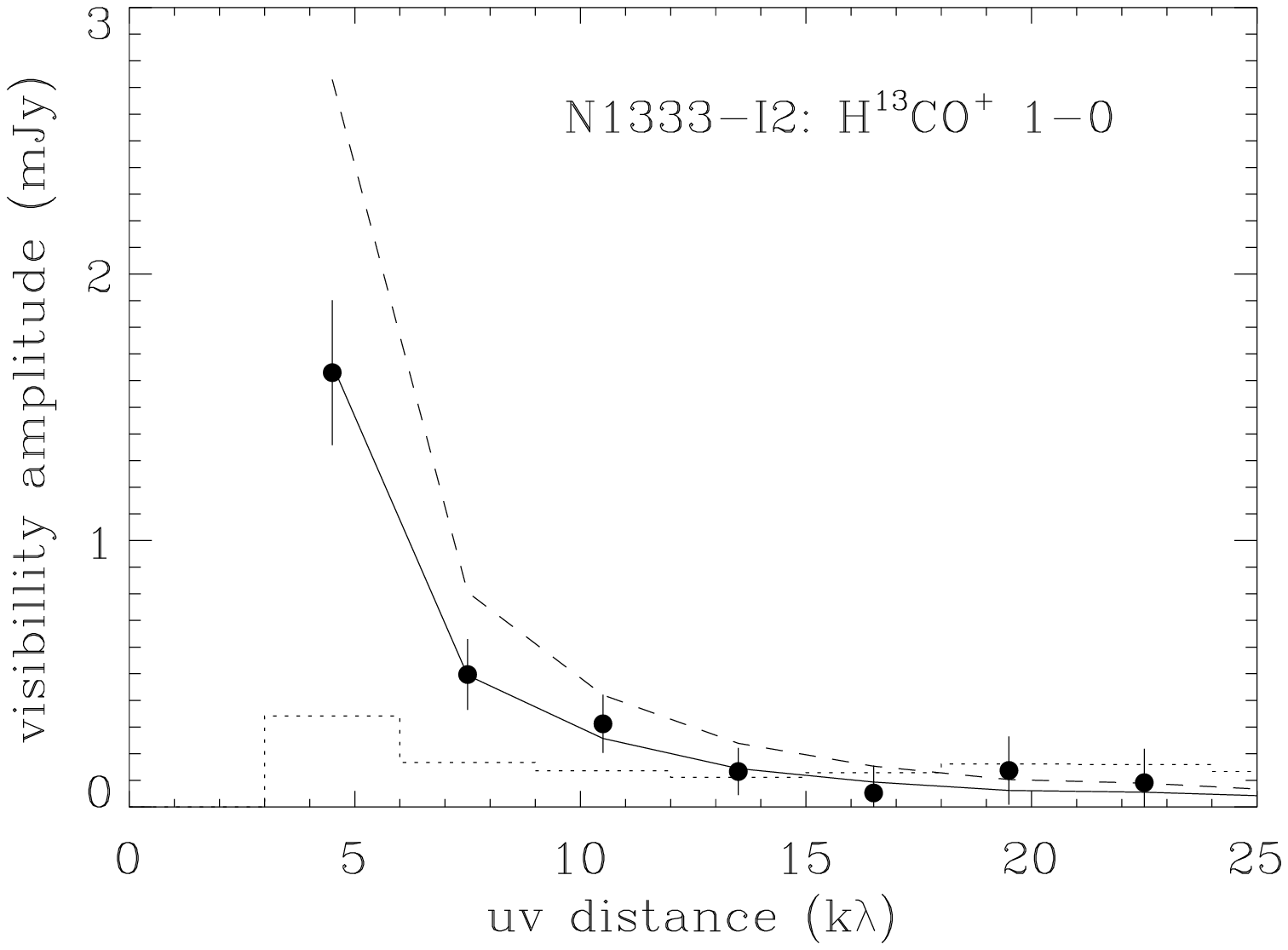}}
\caption{As in Fig.~\ref{cfs_interferometry}, but for \htcop\ emission
as traced by single-dish observations from the Onsala 20m telescope
(44\arcsec\ beam) and interferometry with the OVRO millimeter
array. The \htcop\ interferometer spectra have been scaled by a factor
3 in order to be able to visualize them together with the single-dish
observations. The solid line in both panels indicate the model with
\htcop\ abundance constrained by the high $J$ line observations. The
dashed line indicate the model constrained by the \htcop\ $J=1-0$
line. In the lower panel models both with a turbulent broadening of
0.5 and 1.5~\kms\ have been shown for the abundance constrained by the
$1-0$ line. The integrated intensities are the same,
however.}\label{htcop_interferometry}
\end{figure}

For \htcop\ the situation is a bit more complex. The modeling of the
single-dish lines reveal a picture similar to that of the CO isotopic
species in \citetalias{jorgensen02}; while a constant fractional
abundance can describe the line intensities of higher $J$ lines, the
intensity of the $J=1-0$ line is underestimated by the model. In
\citetalias{jorgensen02} it was suggested that this was due to ambient
cloud material being picked up by the larger single-dish beam. The
same problem may be an issue for \htcop. Fitting the \htcop\ $J=1-0$
single-dish line alone gives an abundance of 8.0\tpt{-11}. With this
abundance, the model can describe the intensity of a spectrum
convolved with a beam equivalent to the single-dish observations as
illustrated in the upper panel of Fig.~\ref{htcop_interferometry}. The
model, however, cannot fit the line profiles simultaneously for the
single-dish and interferometry spectra when a constant turbulent
broadening is adopted. This is likely caused by the larger single-dish
beam picking up the more extended cloud where the velocity
distribution may be different. We emphasize, however, that this is a
significantly smaller effect than what is seen for, e.g., CS, HCN and
HCO$^+$ (Fig.~\ref{resolved_out}). It is also seen that the model
cannot describe the emission on smaller scales when plotting the
visibilities vs. projected baseline length, as illustrated in the
lower panel of Fig.~\ref{htcop_interferometry}. If the lower \htcop\
abundance found from fitting the higher $J$ lines is adopted, the
model perfectly reproduces the observed \htcop\ emission
distribution. This discrepancy suggests that the single-dish beam of
44\arcsec\ picks up the ambient cloud and the envelope around
\object{IRAS2B} as it also is seen in the \htcop\ 1--0 interferometry
maps. This will contribute to the spectrum extracted from the
interferometry cube when convolved with the single-dish beam (e.g. the
upper panel of Fig.~\ref{htcop_interferometry}). The good fit to the
visibility curve in the lower panel of Fig.~\ref{htcop_interferometry}
indicates that the abundance constrained by the higher excitation
single-dish line observations of \htcop\ is representative of the
actual envelope abundance.

\begin{figure}
\resizebox{\hsize}{!}{\includegraphics{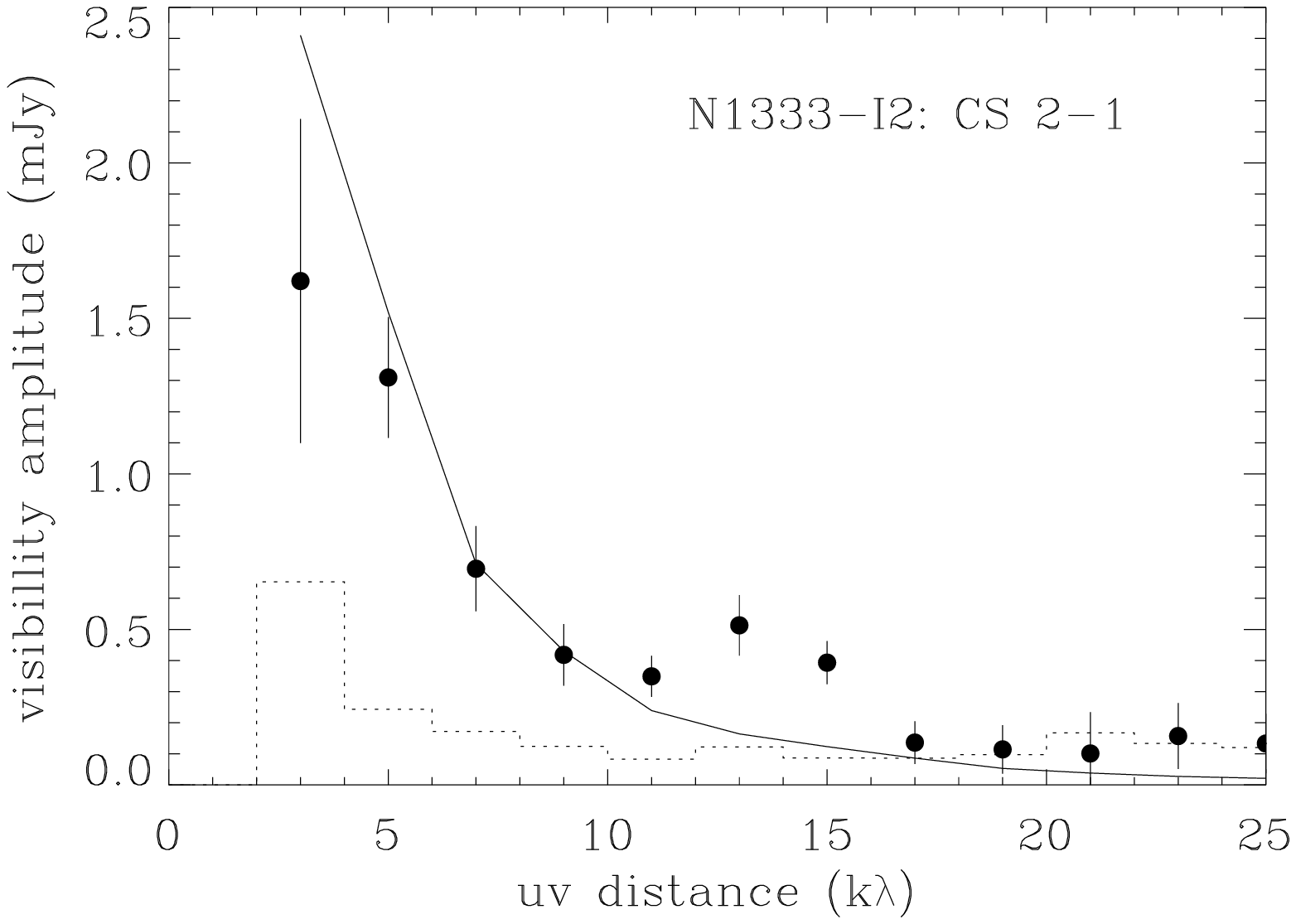}}
\caption{Comparison between interferometry observations and envelope
model for CS.}\label{cs_interferometry}
\end{figure}

It is not possible to account for the observed CS, HCN and HCO$^+$
emission within the envelope models. As can be seen in
Fig.~\ref{cs_interferometry}, the CS line intensity is, for example
reproduced only at intermediate baselines where also the single-dish
line observations are sensitive. On longer baselines the model clearly
breaks down and underestimates the observed emission. It is also
evident that the pronounced double peak structure seen in
interferometry spectra cannot be explained with a simple collapse
model alone. This problem will be further explored in the next
section.

\section{Velocity structure beyond the envelope}\label{velgradient}
\begin{figure*}
\resizebox{\hsize}{!}{\includegraphics{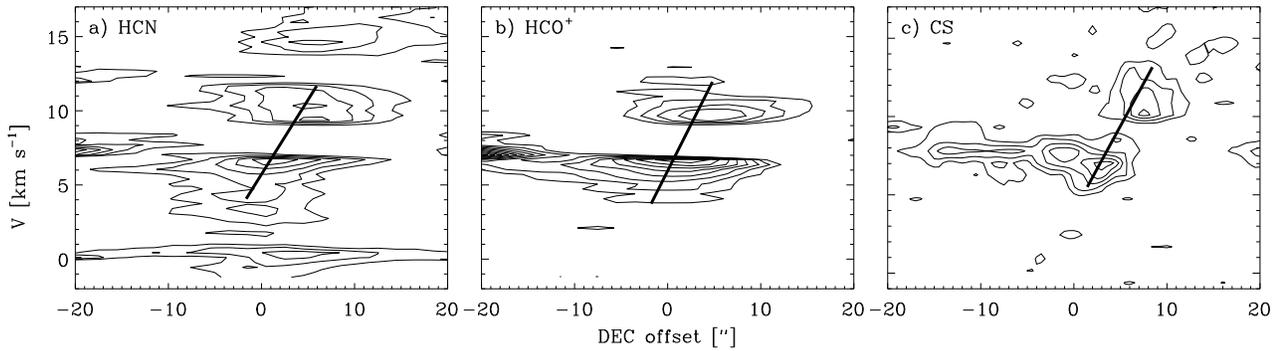}}
\caption{Position-velocity diagrams for a) HCN, b) HCO$^+$ and c)
CS. The solid line indicates a linear gradient fitted to the
centroids for the velocity channel. The hyperfine splitting of HCN is
seen as the extension of emission along the velocity
axis.}\label{pvdia}
\end{figure*}

Fig.~\ref{pvdia} shows position-velocity diagrams of HCO$^+$, HCN, and
CS along the north-south velocity gradient apparent in the
material within $\approx 20$\arcsec\ from IRAS2A as seen in
Fig.~\ref{bima_moment}-\ref{first_order}. The previous section found
that the velocity structure in these lines could not be explained by
the envelope model with infall. Fitting a linear gradient to the
velocity centroid at each offset yields values as given in
Table~\ref{posveltab} for the three species. It is seen that the
fitted velocity gradients agree well and correspond to a weighted
average of $1.10\pm 0.23$ km~s$^{-1}$~arcsec$^{-1}$. This gradient may
be an overestimate of the actual gradient, because the interferometer
predominantly recovers extreme velocities due to resolving out. On the
other hand, velocity gradients inferred from single dish observations
only are biased toward velocities closer to the rest velocity of the
cloud due to the larger beam.

\begin{table}
\caption{Velocity gradient derived through a linear fit to the
centroid of the brightness for each velocity channel as indicated in
Fig.~\ref{pvdia}.}\label{posveltab}
\begin{tabular}{lll}\hline\hline
Molecule & Velocity gradient \\
         & [\kms\ arcsec$^{-1}$]\\ \hline
CS       & $1.09\pm 0.29$    \\
HCN      & $1.00\pm 0.49$    \\
\hcop\   & $1.25\pm 0.53$    \\ \hline
\end{tabular}
\end{table}

The inferred north-south velocity gradient of
1.10~km~s$^{-1}$~arcsec$^{-1}$, or 1.03\tpt{3}~km~s$^{-1}$~pc$^{-1}$,
is two orders of magnitude larger than that inferred by
\cite{wardthompson01} from single-dish CS and HCO$^+$
observations. This increase in velocity gradient from the spatial
scales of the single-dish to the interferometer data is too large to
be explained by differential rotation in either a Keplerian structure
(expected increase of a factor $\sqrt{10}$) or a magnetically braked
core \citep{basu98} (expected increase of a factor 10). Even if the
gradients on single-dish and interferometer scales are unrelated,
Keplerian rotation cannot explain our observed velocities since it
requires an unrealistically large central mass of 31~$M_\odot$.

An alternative explanation for the north-south velocity gradient
around \object{IRAS2A} is that it is part of the north-south outflow detected
on larger scales in CO \citep{liseau88, engargiola99,
knee00}. \cite{engargiola99} concluded that the origin of this flow
lies within a few arcsec from 2A, while \cite{hodapp95} infer a
north-south jet that passes within a few arcsec from 2A from H$_2$
images. Neither this Paper nor \cite{looney00} find evidence for
continuum emission from a second source, although it could be below
the detection limit or be unresolved ($<0.3\arcsec=65$~AU). The
different levels to which HCO$^+$, HCN, and CS trace the east-west and
north-south flows may reflect differences in shock chemistry as the
flows progress through the inhomogeneous cloud environment of
IRAS2. This also serves as caution in interpreting differences in the
spatial extent of, e.g., CO line wing emission as differences in
`dynamic time scales'; this presupposes similar environments in which
the flows propagate.

Instead of two perpendicular outflows, a single, wide-angle
($90^\circ$), northwest--southeast flow could also explain the
observed gradients. In this interpretation, what appear to be two
independent flows actually trace the interaction of the wide-angle
flow with the surrounding material along the sides of the cavity. This
scenario is reminiscent of the wide-angle outflow of \object{B5-IRS1}
\citep{velusamy98}. However, in this scenario the jet-like morphology
of the shocked region east of \object{IRAS2}
\citep{blake96,bachiller98,i2art} is difficult to explain.

\section{Conclusion\label{conclusion}}
This Paper has shown that the envelope model derived from
submillimeter continuum imaging with SCUBA provides a useful framework
to interpret interferometric measurements of continuum and line
emission. It allows separation of small-scale structures associated
with the envelope from small-scale structure in additional elements of
the protostellar environment, such as outflows and disks. Our main
findings are as follows.

\begin{enumerate}
\item{Compact 3~mm continuum emission is associated with the two
protostellar sources NGC~1333-IRAS2A and 2B; the starless core 2C is
not detected, indicating it lacks sufficient central concentration.}
\item{The 3~mm continuum emission around 2A in the interferometer data
is consistent with the extrapolation of the envelope density and
temperature distribution to small scales. Changes in the extent of the
envelope and inclusion of the interstellar radiation field do not
change this conclusion. A density structure as predicted from an
inside-out collapse \citep{shu77} fits the data equally well.}
\item{The 3~mm continuum data show the presence of a 22~mJy unresolved
source, presumably a circumstellar disk of total mass $\gtrsim
0.3$~$M_\odot$.}
\item{Line emission in the optically thin tracers H$^{13}$CO$^+$ and
\cfs\ is also consistent with the extrapolated envelope model. Since
these tracers are optically thin, this suggest that the bulk of the
material is well described by the envelope model.}
\item{Optically thick emission lines of CS, HCO$^+$, and HCN only
trace a small fraction of the material at velocities red- and
blue-shifted by several km~s$^{-1}$. Emission closer to systemic is
obscured by resolved-out large-scale material. The detected emission
is closely associated with two perpendicular outflows directed
east-west and north-south. This suggests that the source 2A is an
unresolved ($<65$ AU) binary.}
\item{The morphology of the line emission in the maps shows that
chemical effects are present. An example is the emission of N$_2$H$^+$
that traces cold material around 2A and that is especially strong
toward the starless core 2C. The emission avoids the region around 2B
and the outflows. We suggest that the dearth of N$_2$H$^+$ emission is
due to destruction through reaction with CO released from ice mantles
in warmed-up regions. This indicates an evolutionary ordering
2C--2A--2B, in order of increasing thermal processing of the
material.}
\end{enumerate}

This work suggests that successful interpretation of the small-scale
structure around embedded protostars requires a solid framework for
the structure of the surrounding envelope on larger scales. In this
framework one can effectively fill in the larger-scale emission that
is resolved out by interferometer observations. The
submillimeter-continuum imaging by instruments like SCUBA has proved
particularly powerful because it does not suffer from chemical effects
that make line emission measurements so complex. On the other hand,
this very chemistry reflects which physical processes are occurring:
e.g., the N$_2$H$^+$ emission that shows the thermal history of the
material.

The success of the envelope model in describing the optically thin
species, such as C$^{34}$S and \htcop\ makes \object{IRAS2} a
promising candidate in order to study the relation between the
envelope chemistry and the spatial distribution of molecular
species. In particular, studies of a larger sample of optically thin
molecular lines at arcsecond scale resolution may probe differences in
the radial distributions of molecules reflecting the
chemistry. \object{IRAS2A} is for this purpose a promising target due
to the relative simplicity of the central envelope component. High
angular resolution, high sensitivity maps may also allow for a more
detailed comparison to models for the protostellar collapse in order
to possibly address the evolution of low-mass protostars in the
earliest stages.

\begin{acknowledgement}
The authors thank Kees Dullemond for use of the CGPLUS program and
discussions of disk models. The research of JKJ is funded by the
Netherlands Research School for Astronomy (NOVA) through a network 2
Ph.D. stipend and research in astrochemistry in Leiden is supported by
a Spinoza grant. This paper made use of data from a range of
telescopes among them the Owens Valley Radio Observatory and
Berkeley-Illinois-Maryland-Association millimeter arrays, Onsala Space
Observatory 20~m telescope and the James Clerk Maxwell Telescope. The
authors are grateful to the staff at all these facilities and their
host institutions for technical support, discussions, and hospitality
during numerous visits.

\end{acknowledgement}

\bibliographystyle{aa}

\end{document}